\documentclass{pasa}

\title[Quasar Broad Emission Line Regions Disk-wind Model]{The Kinematics of Quasar Broad Emission Line Regions Using a Disk-wind Model}
\author[S.~Yong et al.]{Suk Yee Yong$^{1,}$\thanks{E-mail: \texttt{syong1@student.unimelb.edu.au}}, Rachel L.~Webster$^{1}$, Anthea L.~King$^{1}$, Nicholas F.~Bate$^{2}$, Matthew J.~O'Dowd$^{3,4,5}$, \and Kathleen Labrie$^{6}$\\
\affil{$^{1}$School of Physics, University of Melbourne, Parkville, VIC 3010, Australia}%
\affil{$^{2}$Institute of Astronomy, University of Cambridge, Madingley Road, Cambridge CB3 0HA, UK}%
\affil{$^{3}$Department of Physics and Astronomy, Lehman College of the CUNY, Bronx, NY 10468, USA}%
\affil{$^{4}$Department of Astrophysics, American Museum of Natural History, Central Park West and 79th Street, NY 10024-5192, USA}%
\affil{$^{5}$The Graduate Center of the City University of New York, 365 Fifth Avenue, New York, NY 10016, USA}
\affil{$^{6}$Gemini Observatory, Hilo, HI 96720, USA}}%
\jid{PASA}
\doi{10.1017/pas.\the\year.xxx}
\jyear{\the\year}

\usepackage{amsmath,amssymb,gensymb}
\usepackage{graphicx,subcaption,xcolor}
\usepackage{hyperref}
\usepackage[authoryear]{natbib}
\bibpunct{(}{)}{;}{a}{}{,}
\newcommand{\subtext}[2]{\ensuremath{#1_{\text{#2}}}} 
\newcommand{\ion}[2]{#1\,{\scshape{#2}}} 
\newcommand{\ergs}{\ifmmode {\rm erg\,s}^{-1} \else erg\,s$^{-1}$\fi} 
\newcommand{\kms}{\ifmmode {\rm km\,s}^{-1} \else km\,s$^{-1}$\fi} 
\newcommand{\diff}{\mathop{}\!\mathrm{d}} 

\begin{document}

\begin{abstract}
The structure and kinematics of the broad line region (BLR) in quasars are still not well established. One popular BLR model is the disk-wind model that offers a geometric unification of a quasar based on the angle of viewing. We construct a simple kinematical disk-wind model with a narrow outflowing wind angle. The model is combined with radiative transfer in the Sobolev, or high velocity, limit. We examine how angle of viewing affects the observed characteristics of the emission line, especially the line widths and velocity offsets. The line profiles exhibit distinct properties depending on the orientation, wind opening angle, and region of the wind where the emission arises.

At low inclination angle (close to face-on), we find the shape of the emission line is asymmetric with narrow width and significantly blueshifted. As the inclination angle increases (close to edge-on), the line profile becomes more symmetric, broader, and less blueshifted. Additionally, lines that arise close to the base of the disk wind, near the accretion disk, tend to be broad and symmetric. The relative increase in blueshift of the emission line with increasing wind vertical distance is larger for polar winds compared with equatorial winds. By considering the optical thickness of the wind, single-peaked line profiles are recovered for the intermediate and equatorial outflowing wind. The model is able to reproduce a faster response in either the red and blue sides of the line profile found in reverberation mapping studies. A quicker response in the red side is achieved in the model with a polar wind and intermediate wind opening angle at low viewing angle. The blue side response is faster for an equatorial wind seen at high inclination.
\end{abstract}


\maketitle

\section{Introduction} \label{sec:intro}

The immense power of a quasar comes from accreting mass onto the central black hole via the accretion disk. During the accretion process, gravitational potential energy is converted into radiation via viscous dissipation, creating the continuum we observe. The emitted continuum radiation photoionises the surrounding gas, which is deep in the potential well of the black hole, allowing the formation of the broad line region (BLR). Since the BLR is too small spatially to be resolved by modern telescopes even for the most nearby objects, its geometry, kinematics, and dynamics remain elusive.

A common feature of a quasar spectrum in the optical and ultraviolet regimes is the broad emission lines (BELs). The properties of BELs provide crucial details on the nature of the BLR. The BEL profiles show wide diversity in their widths and shapes. Their line widths commonly exceed $10^{3}\,\kms$ and can extend to $10^{4}\,\kms$. In general, the width of the high-ionisation lines (HILs), such as \ion{C}{iv}, are also found to be broader than the low-ionisation lines (LILs), such as \ion{Mg}{ii} \citep[e.g.,][]{Shuder:1982,Mathews+Wampler:1985}. The variation in line profile shapes reflects the dynamics of the emitting gas in the BLR. This provides valuable information on the structure and geometry of the emission line region.

One of the techniques used to probe the structure of the BLR is reverberation mapping \citep[RM;][]{Blandford+McKee:1982,Peterson:1993}. This method measures a time delay between emission line flux variations and continuum flux variations. Analyses based on RM results infer that the BLR has a stratified ionisation structure \citep{Peterson+Wandel:1999,Kollatschny:2003,Peterson+:2004}. The HILs are found to have a shorter time lag compared to that of LILs. This suggests that the HILs are located closer to the central engine, while the LILs are situated further out \citep{Gaskell+Sparke:1986,Clavel+:1991,Peterson+Wandel:1999,Kollatschny:2003}.

Variability in the line profile provides a way to extract information on the geometry and kinematics of the BLR \citep[e.g.,][]{Bahcall+:1972,Blandford+McKee:1982,Capriotti+:1982,Horne+:2004,Pancoast+:2011}. By measuring the time delays of emission lines as a function of line-of-sight velocity, a velocity-delay map can be constructed, which makes it possible to predict whether the BLR dynamics are dominated by inflow, outflow, or rotation \citep{Horne+:2004}. Different studies have reached the following conclusions. An asymmetric velocity profile with faster response in the red or blue wing of the line tends to be associated with infall or outflow, respectively. For virialised gas motion, the velocity profile is symmetric with quicker response in the line wings compared to the line core. Through analyses of the recovered velocity-delay maps from velocity-resolved RM, most objects show combinations of infall and rotation since the red side of the line wing tends to have a shorter lag than the blue side \citep[e.g.,][]{Gaskell:1988,Koratkar+Gaskell:1989,Crenshaw+Blackwell:1990,Korista+:1995,Ulrich+Horne:1996,Kollatschny:2003,Bentz+:2010,Grier+:2013}. However, a signature of outflow, with the blue line wing leading, has been found in some objects \citep[e.g., NGC~3227,][]{Denney+:2009}.

Among the various proposed models of the BLR are the discrete cloud and disk-wind models \citep[see reviews by e.g.,][]{Korista:1999,Eracleous:2006}. The discrete cloud model consists of discrete optically thick gas clouds photoionised via the continuum source emission. Although this model is capable of explaining the observed spectral characteristics, it suffers from a few complications. One main issue is the confinement problem: without a way to confine the clouds, they would simply evaporate \citep[see review by e.g.,][]{Mathews+Capriotti:1985}. Several solutions to this problem have been suggested, such as a hot intercloud medium \citep{Krolik+:1981} and magnetic fields \citep{Rees:1987}.

The disk-wind model is composed of a wind that originates close to the accretion disk and is accelerated away from the central black hole due to some driving mechanisms. In this scenario, the line emitting materials are continuously distributed within the optically thin wind and consequently the wind does not encounter the confinement problem. The area of the disk covered by the outflowing wind can be wide, such as \citet{Murray+:1995} model, or narrow as in \citet{Elvis:2000} model. The disk-wind model explains the existence of BELs and broad absorption lines (BALs), and why the BALs are only detected in a small subset of the quasar spectra. The disk-wind model is also able to explain the few percent of radio sources that exhibit broad double-peaked emission lines \citep{Eracleous+Halpern:1994,Eracleous+Halpern:2003,Strateva+:2003}.

It has been known for several decades that there is an offset between different ionisation lines \citep[e.g.,][]{Gaskell:1982,Wilkes:1986,Espey+:1989,Tytler+Fan:1992,McIntosh+:1999,VandenBerk+:2001,Shen+:2016}. The HILs are usually seen blueshifted relative to LILs. This blueshift can be interpreted as a consequence of an outflowing wind component and obscuration by the disk \citep{Gaskell:1982,Leighly:2004,Richards+:2011}, and is one of the major motivators for the disk-wind model.

Currently, the driving mechanism of the wind is still unclear. The three major drivers that have been proposed are thermal or gas pressure \citep{Weymann+:1982,Begelman+:1991}, radiation pressure due to spectral lines or line driving  \citep{Shlosman+:1985,Arav+:1994,Murray+:1995}, and magnetocentrifugal pressure due to the accretion disk \citep{Blandford+Payne:1982,Emmering+:1992,Konigl+Kartje:1994}. Low velocity gas of $\sim 1000\,\kms$ can be explained by the thermally driven wind model. However, to create the more commonly observed high velocity gas, radiation pressure and magnetically driven wind models are more likely mechanisms. In the series of papers by \citet{Murray+:1995,Chiang+Murray:1996,Murray+Chiang:1997}, they demonstrated that an optically thick wind driven by radiative pressure is able to reproduce realistic emission line profiles.

The goal of this paper is to provide qualitative constraints on BEL models through simple kinematical modelling of a basic disk-wind model. We explore the effect of orientation on the shapes of the emission lines for outflowing winds with narrow opening angles. The details of the line shapes, in particular the line widths and velocity offsets, will hopefully enable us to describe the kinematics and dynamics of the BLR. The overview of the paper is as follows. The details on modelling the disk-wind are explained in Section~\ref{sec:modelwind}. Section~\ref{sec:elprofiles} presents the generated line profiles. In Section~\ref{sec:discussion}, the implications of wind opening angle, inclination angle, and wind region on the line widths and velocity offsets are explored. The conclusions are given in Section~\ref{sec:summary}.

\section{Modelling the Wind} \label{sec:modelwind}

\begin{figure}
  \centering
  \includegraphics[width=\columnwidth]{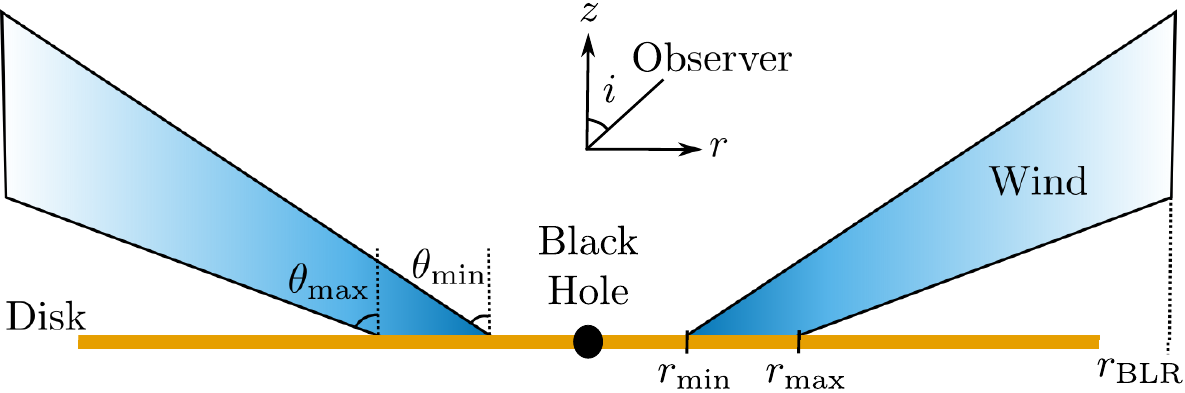}
  \caption{A sketch of the geometry and main parameters of our cylindrical disk-wind model.}
  \label{fig:cywindmodel}
\end{figure}

The kinematics of our BLR disk-wind model is adopted from \citet{Shlosman+Vitello:1993}, designed initially for cataclysmic variables (CVs). It has since been extended to model quasars \citep{Higginbottom+:2013,Higginbottom+:2014,Matthews+:2016,Yong+:2016}. A schematic of the model is depicted in Fig.~\ref{fig:cywindmodel}. The geometry of the wind is similar to the models proposed by \citet{Murray+:1995,Elvis:2000,Elvis:2004} to describe the phenomenology of BELs and BALs in quasars. In our model, the vertical component of the wind proposed in \citet{Elvis:2000,Elvis:2004}, where the wind is lifted vertically off the disk before being accelerated outwards, is not incorporated. We also incorporate radiative transfer effects to generate the line profiles using the Sobolev (high velocity gradient) approximation, following previous work \citep{Chiang+Murray:1996,Murray+Chiang:1997,Flohic+:2012}.

\subsection{Wind Kinematics} \label{ssec:kinematics}

The details of the wind geometry and kinematics are as described in \citet{Yong+:2016}. However, we implement several modifications to the creation of the line profile. We briefly review the kinematics of the outflow and highlight any updates that we made.

The BLR model is assumed to have an axially symmetric geometry, which we describe using cylindrical coordinates $(r, \phi, z)$. The radial and azimuthal coordinates, $r$ and $\phi$, are located on the $xy$-plane, which is defined as the plane of the accretion disk surface. The $z$-axis is defined as the rotation axis of the disk. The angle between the $z$-axis and the line-of-sight of the observer is defined by the inclination angle, $i$.

The disk-wind model is described by two main components: the accretion disk and wind. The accretion disk is assumed to be flat and geometrically thin, but optically thick such that the far side of the disk is obscured. A conical outflowing wind emanates from the accretion disk with inner and outer radii of \subtext{r}{min} and \subtext{r}{max}. The boundary of the wind opening angle is between \subtext{\theta}{min} and \subtext{\theta}{max}. The wind streamline spirals upwards in a three-dimensional helical motion at a constant opening angle, $\theta$. The opening angle of the streamlines are distributed linearly with respect to the radius of the streamline origin, which is set by $\gamma=1$.

The total velocity at any given point inside the wind is expressed in terms of the poloidal velocity, $v_{l}$, and rotational velocity, $v_{\phi}$. The poloidal velocity describes the velocity component along the streamline, which consists of a combination of the radial component, $v_{r}$, and vertical component, $v_{z}$. It specifies the velocity in the $rz$-plane, which is given by
\begin{equation}
v_{l}=v_{0}+(v_{\infty}-v_{0})\left[\frac{(l/R_{v})^{\alpha}}{(l/R_{v})^{\alpha}+1}\right],
\label{eqn:poloidalvelocity}
\end{equation}
where $v_{0}$ is the initial poloidal velocity at the base of the disk and $l$ is the poloidal distance along a streamline. The acceleration scale height, $R_{v}$, describes the point at which the wind achieves half of its terminal velocity, $v_{\infty}$. The power law index that adjusts the acceleration of the wind, $\alpha$, is taken to be 1, such that the acceleration increases slowly with increasing poloidal distance. The terminal velocity is set to be equal to the escape velocity, \subtext{v}{esc}. The initial rotational velocity on the surface of the disk is presumed to be Keplerian, $v_{\phi,0}=(G\subtext{M}{BH}/r_{0})^{1/2}$. The assumption that the wind conserves specific angular momentum implies that the rotational velocity decreases linearly as the wind is accelerated radially outwards.

For a particular point $(r,z)$ in the wind, the density, $\rho$ is determined using the mass continuity equation
\begin{equation}
\rho(r,z)=\frac{r_{0}}{r}\frac{\diff r_{0}}{\diff r}\frac{\dot{m}(r_{0})}{v_{z}(r,z)},
\label{eqn:density}
\end{equation}
where $\dot{m}$ is the local mass-loss rate per unit surface of the disk,
\begin{align}
\dot{m}(r_{0}) \propto \subtext{\dot{M}}{wind}r^{\lambda}_{0}\cos\theta(r_{0}).
\label{eqn:masslossrate}
\end{align}
Assuming an accretion efficiency of $\eta=0.1$, the total mass accretion rate for a source with high luminosity of $L \approx 10^{46}\,\ergs$ and black hole mass of $10^{8}\,M_{\odot}$ is $\subtext{\dot{M}}{acc} \approx 2\,M_{\odot}\,$yr$^{-1}$ \citep{Peterson:1997}. The total mass-loss rate of the wind, \subtext{\dot{M}}{wind}, is fixed to be equal to the total mass accretion rate, $\subtext{\dot{M}}{acc}=2\,M_{\odot}\,$y$^{-1}$. The mass-loss rate exponential, $\lambda$, is taken to be 0, indicating a uniform mass-loss with radius.

The parameter values used in this paper are the same as those chosen for \citet{Elvis:2004} model in \citet{Yong+:2016}, except the wind opening angle. The list of parameters in the model is shown in Table~\ref{tab:parametersmodel}. The size of the BLR wind region is bounded within $\subtext{r}{BLR}=2 \times 10^{17}\,$cm, in both radius and height. The value is chosen such that the effects of the poloidal velocity can be seen. This fiducial radius generally agrees with results from RM and microlensing. For a quasar with black hole mass of $\subtext{M}{BH} \sim 10^{8}\,M_{\odot}$, RM studies found that the size of the BLR using Balmer lines is about $1 \times 10^{17} \text{--} 5 \times 10^{17}\,$cm \citep{Wandel+:1999,Kaspi+:2000}. Microlensing measurements of the quasar QSO 2237+0305 estimates the BLR radius for HIL \ion{C}{iv} to be $\sim 2 \times 10^{17}\,$cm with $\subtext{M}{BH} \sim 10^{8.3}\,M_{\odot}$ \citep{Sluse+:2011}. For $\subtext{M}{BH} \sim 4 \times 10^{8}\,M_{\odot}$ BAL quasar H1413+117, the BLR size is $\gtrsim 2.9 \times 10^{16}\,$cm \citep{ODowd+:2015}.

\begin{figure}
  \centering
  \includegraphics[width=0.8\columnwidth]{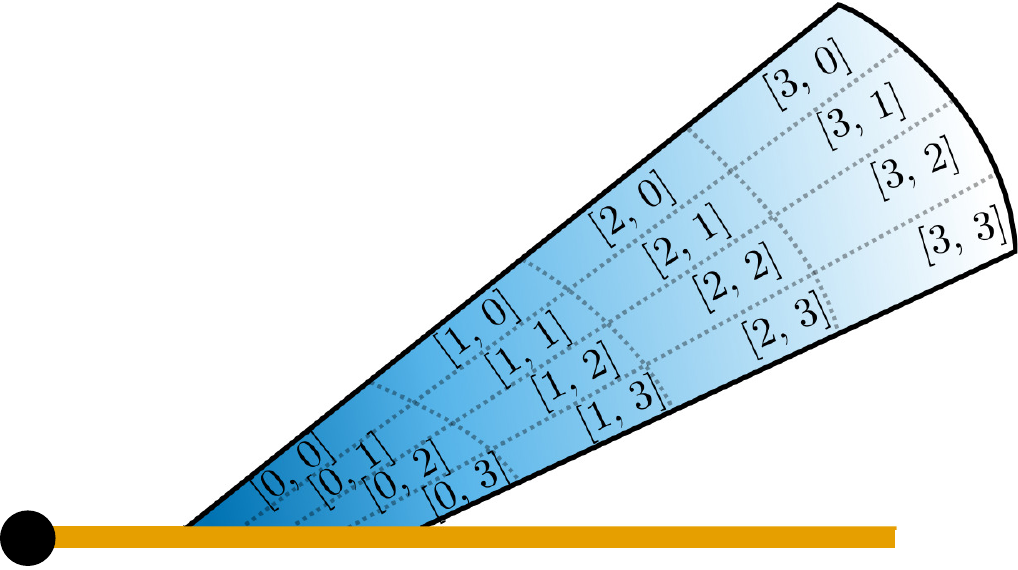}
  \caption{The `wind zones' are numbered by rows and columns, from bottom to top and left to right.}
  \label{fig:cywindzone}
\end{figure}

Based on the evidence that the fraction of BAL quasars is around 20\% of the overall quasar population (\citealt{Weymann+:1991,Hewett+Foltz:2003,Knigge+:2008,Allen+:2011}; though note the results in \citealt{Yong+:2017}), we set the wind to have a narrow opening angle of $10\degree$. However, due to the anisotropic continuum radiation from the accretion disk, the fraction of BALs in optical flux-limited samples might be larger when the outflowing wind opening angle is close to equatorial \citep{Krolik+Voit:1998}. Scattering attenuation of the continuum may also induce substantial bias on the true BAL covering fraction \citep{Goodrich:1997}. These factors are ignored for simplicity. We test ranges of minimum and maximum narrow wind opening angles from polar ($\subtext{\theta}{min}=5\degree; \subtext{\theta}{max}=15\degree$), intermediate (e.g., $\subtext{\theta}{min}=40\degree; \subtext{\theta}{max}=50\degree$), to equatorial ($\subtext{\theta}{min}=75\degree; \subtext{\theta}{max}=85\degree$). To mimic the stratified structure of the wind, we separate the wind into `wind zones' of 4 rows and 4 columns that consist of narrow streams of cones, as illustrated in Fig.~\ref{fig:cywindzone}. An individual zone is designated by the location of its row and column $[a, b]$, starting from the base of the wind, closest to the central ionising source, to increasing radial distance. Quantitative effects due to different choices of some parameter values will be explored in Section~\ref{ssec:paramsensitivity}.

\begin{table*}
\caption{Adopted parameter values in the fiducial model. The values that are different from those used in \citet{Yong+:2016} are marked with an asterisk.}
\label{tab:parametersmodel}
\begin{center}
  \begin{tabular*}{\textwidth}{@{}l\x l\x c@{}}
  \hline\hline
  Parameter & Notation & Value \\
  \hline
  Black hole mass & \subtext{M}{BH} ($10^{8}\,M_{\odot}$) & 1.0 \\
  Wind radius & \subtext{r}{min}; \subtext{r}{max} ($10^{16}\,$cm) & 1.0; 2.0 \\
  Wind opening angle & \subtext{\theta}{min}; \subtext{\theta}{max} & Within $10\degree$ $^{*}$ \\
  Concentration of streamline & $\gamma$ & 1.0 \\
  Initial poloidal velocity & $v_{0}$ (\kms) & 6.0 \\
  Scale height & $R_{v}$ ($10^{16}\,$cm) & 25.0 \\
  Acceleration power law index & $\alpha$ & 1.0 \\
  Mass-loss rate exponent & $\lambda$ & 0.0 \\
  Total mass-loss rate & \subtext{\dot{M}}{wind} ($M_{\odot}\,$yr$^{-1}$) & 2.0 \\
  \hline\hline
  \end{tabular*}
\end{center}
\end{table*}

\subsection{Disk-wind Radiative Transfer} \label{ssec:radtransfer}

When the flow speed is much larger than the mean thermal speed of the gas, radiation at a given frequency as seen by a fixed observer comes primarily from a unique mathematical surface (iso-velocity surface). In such situations, the full radiative transfer can be treated by the approximate escape probability method or Sobolev approximation. In this case, the wind opacity in a given direction mostly depends on the velocity gradient in that direction. Following \citet{Rybicki+Hummer:1978,Rybicki+Hummer:1983} prescriptions, the monochromatic specific luminosity, $\mathcal{L}_{\nu}$, with frequency $\nu$ in the direction $\hat{n}$ over volume $V$ is given by
\begin{align}
\mathcal{L}^{\mathrm{S}}_{\nu}(\hat{n}) &= \int I_{\nu} \diff V \nonumber \\
&= \int k(r_{s})S_{\nu}(r_{s}) \frac{1-e^{\tau}}{\tau} \delta\left[\nu - \nu_{0}\left(1+\frac{\subtext{v}{los}}{c}\right)\right] \diff V,
\label{eqn:msluminosity_sob}
\end{align}
where $I_{\nu}$ is the intensity. Assuming that the line-of-sight does not intersect multiple resonant surfaces, which occurs when $\subtext{v}{los}/c=(\nu-\nu_{0})/\nu_{0}$, where \subtext{v}{los} is the line-of-sight velocity of the particle and $\nu_{0}$ is the central frequency. The integrated line opacity, $k(r_{s})$, and source function, $S(r_{s})$, are written as functions of spherical radius. We have taken a fiducial $S'(r_{s})=k(r_{s})S(r_{s}) \propto r_{s}^{-\beta}$ as a power law function with exponent, $\beta \simeq 1$.

Within the Sobolev approximation, the optical depth is
\begin{align}
\tau=\frac{\xi}{|Q|},
\label{eqn:opticaldepth}
\end{align}
where the $\xi=kc/\nu_{0}$. To investigate the effects of optical depth, we choose to investigate two cases, one where $\xi$ is $1\,$s$^{-1}$ and another where $\xi$ is $10^{10}\,$s$^{-1}$. These values were chosen so that the optical depth of all the points in the wind are optically thin, $\tau<1$, in the first case, and optically thick for the other. The parameter $Q \equiv \hat{n} \cdot \mathbf{\Lambda} \cdot \hat{n}$ is the double-dot product of the strain tensor, $\mathbf{\Lambda}$, with the line-of-sight vector, $\hat{n}$, and describes the gradient along the line-of-sight wind velocity. In cylindrical coordinates, $Q$ is expressed as
\begin{align}
Q &= \sin^{2} i[\Lambda_{rr}\cos^{2}\phi + \Lambda_{\phi\phi}\sin^{2}\phi - 2\Lambda_{r\phi}\sin\phi\cos\phi] \nonumber \\
& \quad \cos i[2\Lambda_{rz}\sin i\cos\phi + \Lambda_{zz}\cos i - 2\Lambda_{\phi z}\sin i\sin\phi].
\label{eqn:Qnln}
\end{align}
Because of azimuthal symmetry, the contribution from $\partial/\partial\phi=0$ and the elements of the strain tensor \citep{Chajet+Hall:2013} are as follows
\begin{align}
\Lambda_{r\phi}=\frac{1}{2}\left(\frac{\partial v_{\phi}}{\partial r} - \frac{v_{\phi}}{r}\right), \quad
\Lambda_{rz}=\frac{1}{2}\left(\frac{\partial v_{r}}{\partial z} + \frac{\partial v_{z}}{\partial r}\right), \nonumber \\
\Lambda_{\phi z}=\frac{1}{2}\frac{\partial v_{\phi}}{\partial z}, \quad
\Lambda_{rr}=\frac{\partial v_{r}}{\partial r}, \quad
\Lambda_{\phi\phi}=\frac{v_{r}}{r}, \quad
\Lambda_{zz}=\frac{\partial v_{z}}{\partial z}.
\end{align}

\subsection{Emission Line Construction} \label{ssec:elconstruction}

Since this study is focussed on understanding the kinematical signatures of BELs, no attempt has been made to include the effects of photoionisation. After the kinematical disk-wind model is initialised, the code proceeds by populating particles within the boundaries of each `wind zone'. A Monte Carlo simulation is implemented to generate a large number of particles. Random points are first created in spherical coordinates $(l, \phi, \theta)$ such that the wind forms an angle $\theta$ at a particular $l$ in the `wind zone'. The coordinates are then transformed to cylindrical coordinates $(r, \phi, z)$ in order to evaluate the projected velocity along the line-of-sight, \subtext{v}{los}. For a given zone, the \subtext{v}{los} is computed for a range of inclination angles, $i$, from $5\degree$ to $85\degree$, and binned into histograms. The counts in each bin are weighted by a density distribution and intensity from radiative transfer. To produce a smooth line profile, the distribution is also convolved using a Gaussian kernel with standard deviation of 3. The time lag of each particle due to the light travel time is determined from the centre of the ionising source. The line profile is separated into the blue and red sides from the median velocity, and the mean time delays, $\langle\tau\rangle$, are calculated for both sides. The difference in mean time delay between the blue and the red side, $\langle\subtext{\tau}{b}\rangle-\langle\subtext{\tau}{r}\rangle$, is then calculated. These steps are repeated for every `wind zone'.

The resulting line profiles are purely based on the kinematics of the wind with optical depth correction. Photoionisation and line driving mechanisms are excluded in our simulations. This should not have a significant effect on the line profiles derived from local sections of the wind, within which incident flux is relatively constant. In addition, plausible clumpiness in the wind \citep{Matthews+:2016} and obscuration by the dusty torus \citep{Krolik+Begelman:1988} are not included.

\section{Emission Line Profiles} \label{sec:elprofiles}

The width and relative velocity shifts of the emission lines are highly dependent on the wind opening angles, inclination, and `wind zone' position. We investigate the effects of changing these parameters on the line properties. Figures~\ref{fig:lpzone_xi0} and \ref{fig:lpzone_xi10} present the generated emission lines as a function of viewing angle, $i=5\degree\text{--}85\degree$, for optically thin wind with $\xi=1\,$s$^{-1}$ and optically thick wind with $\xi=10^{10}\,$s$^{-1}$. Each panel in the `wind zones' represents the location in the wind as defined in Fig.~\ref{fig:cywindzone}. The emission line profiles in some zones exhibit small structures. This is caused by a resolution issue and primarily affects zones with a huge density variation, specifically those near the base of the wind $[0, b]$.

\subsection{Emission Line Shape} \label{ssec:elshape}

In all cases, the line profiles are broadest at the base of the wind $[0, b]$. The line widths also decrease with increasing poloidal distances. However, for intermediate (Figs.~\ref{fig:lpzone_t40-50_xi0} and \ref{fig:lpzone_t40-50_xi10}) and equatorial (Figs.~\ref{fig:lpzone_t75-85_xi0} and \ref{fig:lpzone_t75-85_xi10}) wind opening angles, the widths start to become broader at a point above half of the total poloidal distance, i.e., `wind zones' of $[2, b]$ and $[3, b]$, due to an increase in poloidal velocity.

The emission lines also tend to be more blueshifted, i.e., towards the negative side from the central axis of the line profile, as the wind travels farther away off the base in the direction of increasing height from $[0, b]$ to $[3, b]$. This effect is more prominent for polar outflowing wind (Figs.~\ref{fig:lpzone_t5-15_xi0} and \ref{fig:lpzone_t5-15_xi10}) but is present in all wind models. For more equatorial winds, the relative velocity shift of the emission lines between regions is reduced. The line profiles become more redshifted in the direction of increasing horizontal distance from $[a, 0]$ to $[a, 3]$.

At high inclination angles close to $i=90\degree$ (edge-on), the line profiles are roughly symmetric and less blueshifted relative to the axis centre. The lines are also broader compared to those for face-on viewing angle around $i=0\degree$. When the viewing angle is close to face-on, the lines are asymmetric and exhibit a negative velocity offset. The differences are less pronounced in wind regions near the wind base.

A comparison between Figs.~\ref{fig:lpzone_xi0} and \ref{fig:lpzone_xi10} illustrates the effects of optical depth on the line profiles. In the $\xi=1\,$s$^{-1}$ case, the wind is optically thin, while the $\xi=10^{10}\,$s$^{-1}$ case yields an optically thick wind. In the intermediate and equatorial wind, the double peaks combined to form a single peak line profile for the $\xi=10^{10}\,$s$^{-1}$ case, except for zones nearest to the base. However, the line profiles for the polar wind still exhibit double-peaked features even when the emission is optically thick. This is because the velocity shear $|(\diff v_{l}/\diff r)/(\diff v_{\phi}/\diff r)|$ is low \citep{Chiang+Murray:1996,Murray+Chiang:1997}. Single-peaked lines are expected to form when the radial shear is larger than the Keplerian shear as photons are more likely to escape radially in this case.

\begin{figure*}[!htbp]
\centering
\begin{subfigure}[t]{0.83\textwidth}
  \includegraphics[width=\textwidth,keepaspectratio]{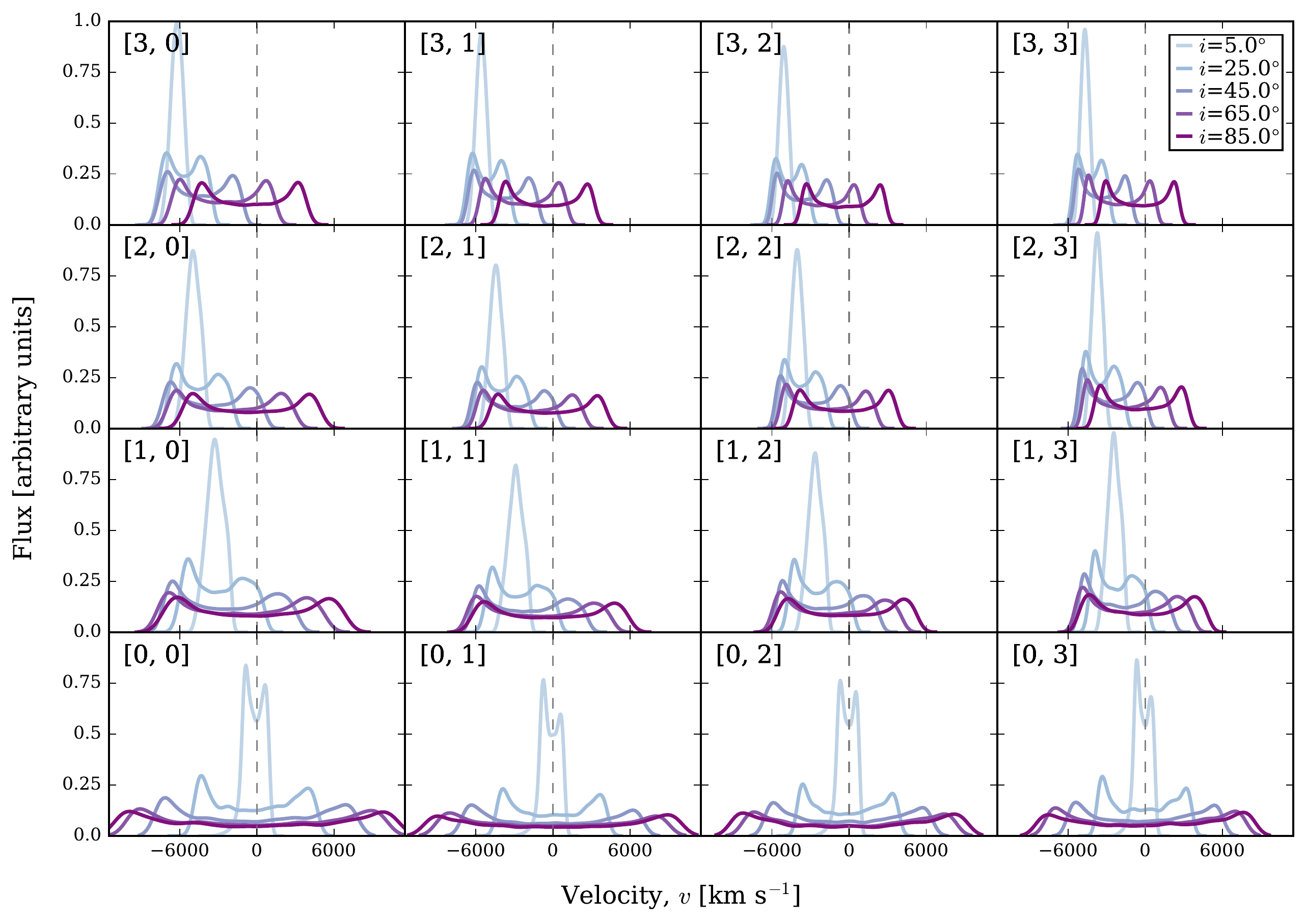}
  \caption{Optically thin polar wind with opening angle of $5\degree \text{--} 15\degree$.}
  \label{fig:lpzone_t5-15_xi0}
\end{subfigure}
\begin{subfigure}[t]{0.83\textwidth}
  \includegraphics[width=\textwidth,keepaspectratio]{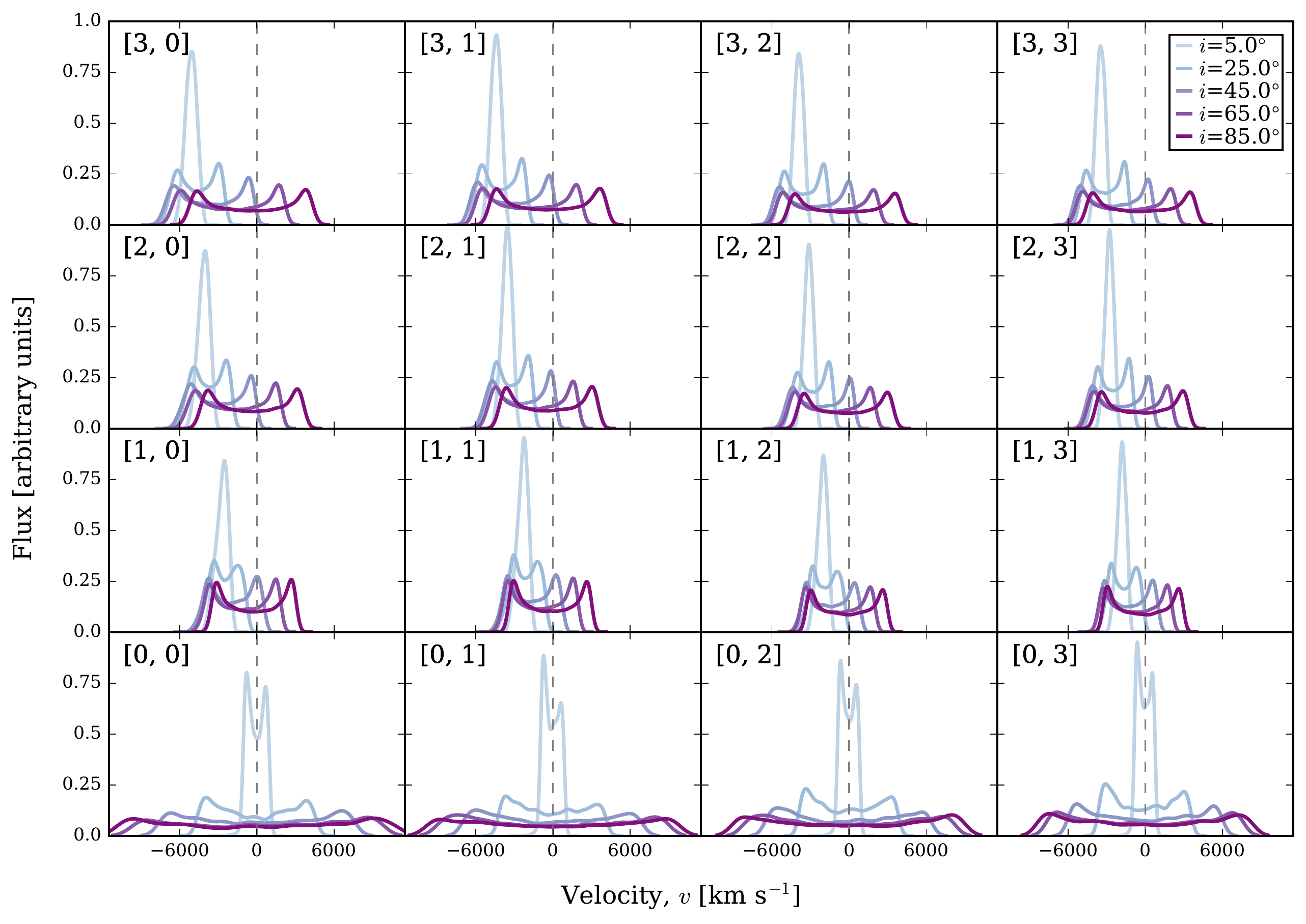}
  \caption{Optically thin intermediate wind with opening angle of $40\degree \text{--} 50\degree$.}
  \label{fig:lpzone_t40-50_xi0}
\end{subfigure}
\caption{Simulated emission line profiles as a function of inclination angle for optically thin wind with $\xi=1\,$s$^{-1}$. The position of the `wind zone' $[a, b]$ is indicated on the top left of each panel. The dashed line shows the centre of the axis. \label{fig:lpzone_xi0}}
\end{figure*}%
\begin{figure*}[!htbp]
\ContinuedFloat
\centering
\begin{subfigure}[t]{0.83\textwidth}
  \includegraphics[width=\textwidth,keepaspectratio]{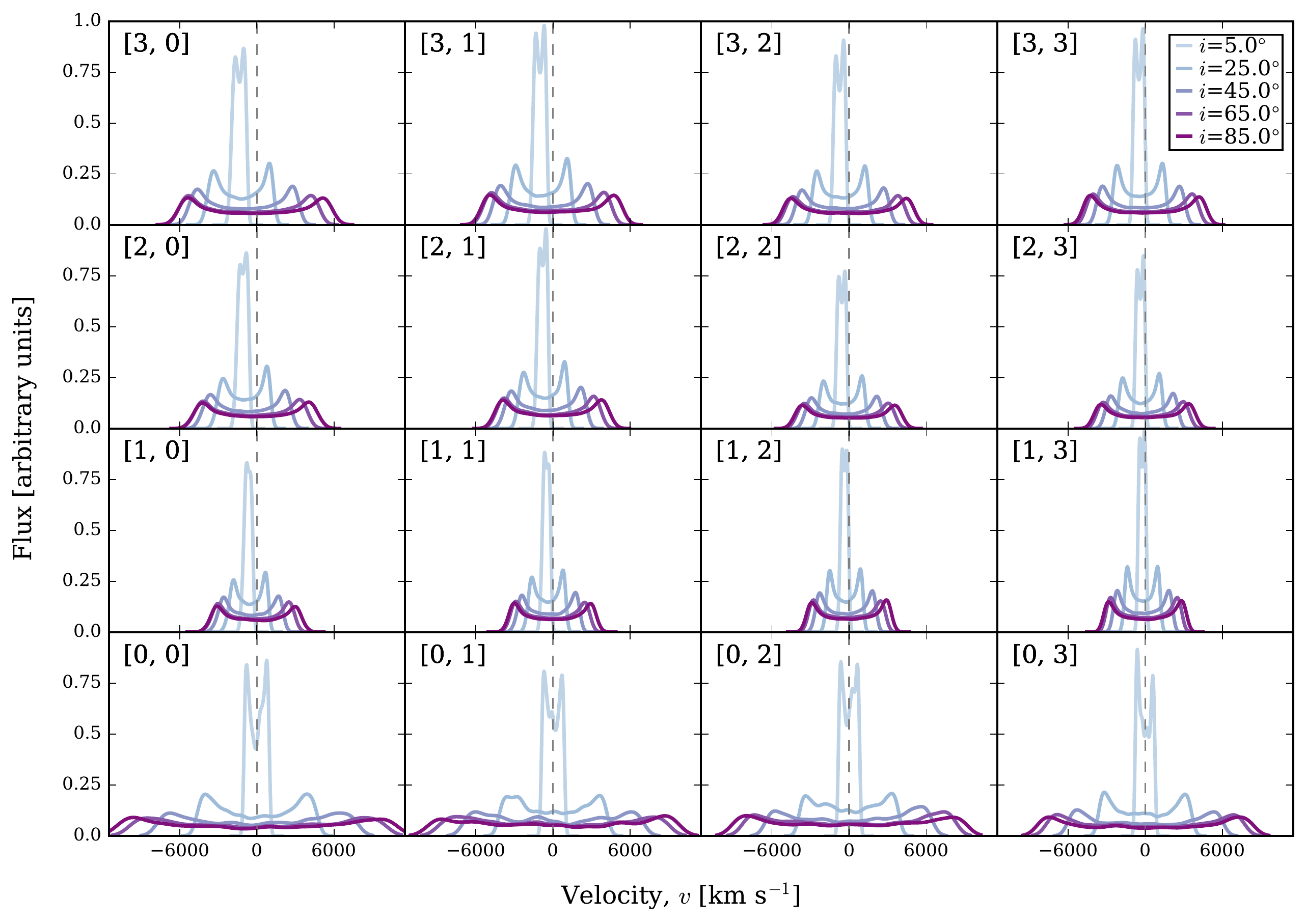}
  \caption{Optically thin equatorial wind with opening angle of $75\degree \text{--} 85\degree$.}
  \label{fig:lpzone_t75-85_xi0}
\end{subfigure}
\caption{--- Continued \label{fig:lpzone_xi0}}
\end{figure*}

\begin{figure*}[!htbp]
\centering
\begin{subfigure}[t]{0.83\textwidth}
  \includegraphics[width=\textwidth,keepaspectratio]{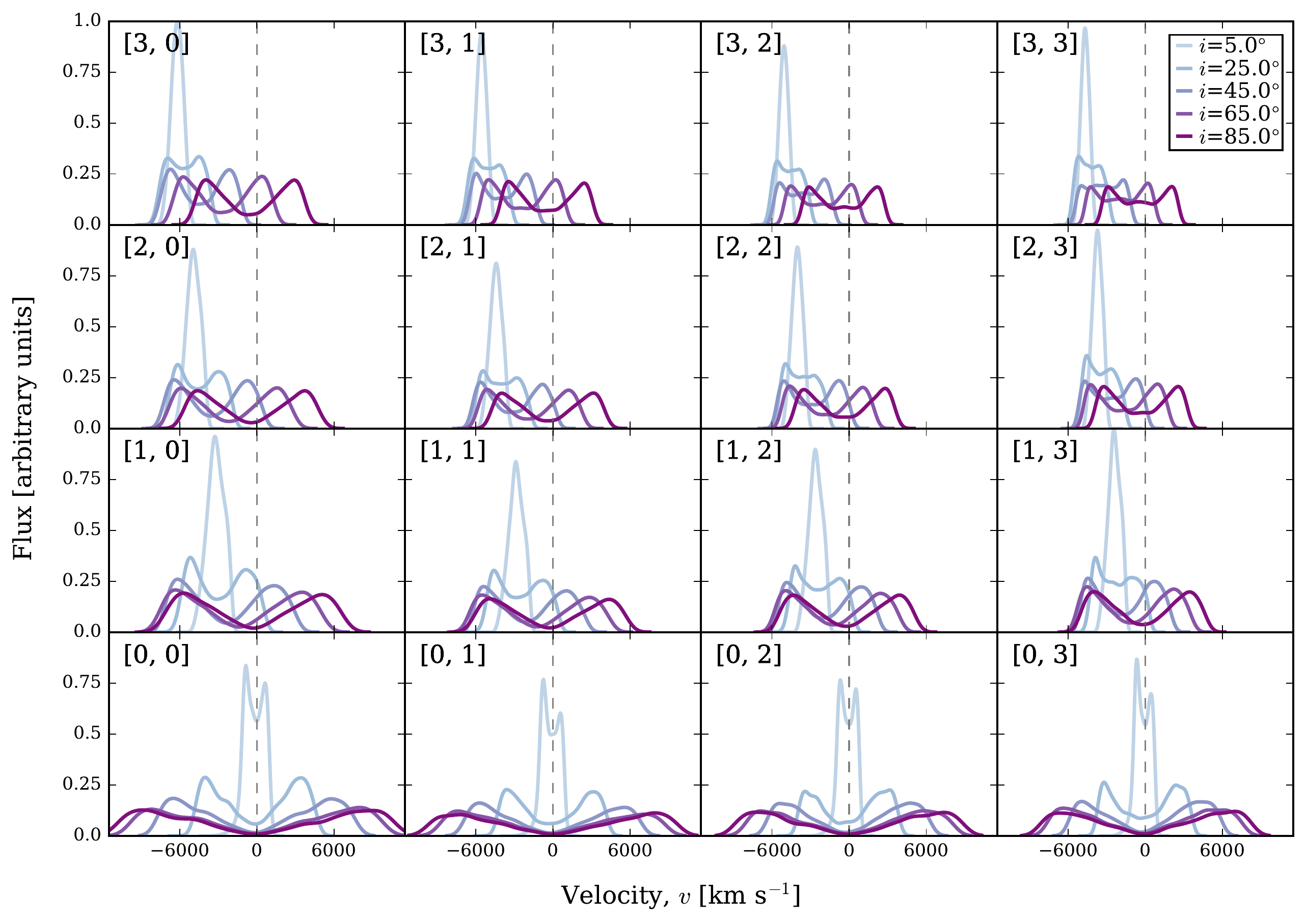}
  \caption{Optically thick polar wind with opening angle of $5\degree \text{--} 15\degree$.}
  \label{fig:lpzone_t5-15_xi10}
\end{subfigure}
\begin{subfigure}[t]{0.83\textwidth}
  \includegraphics[width=\textwidth,keepaspectratio]{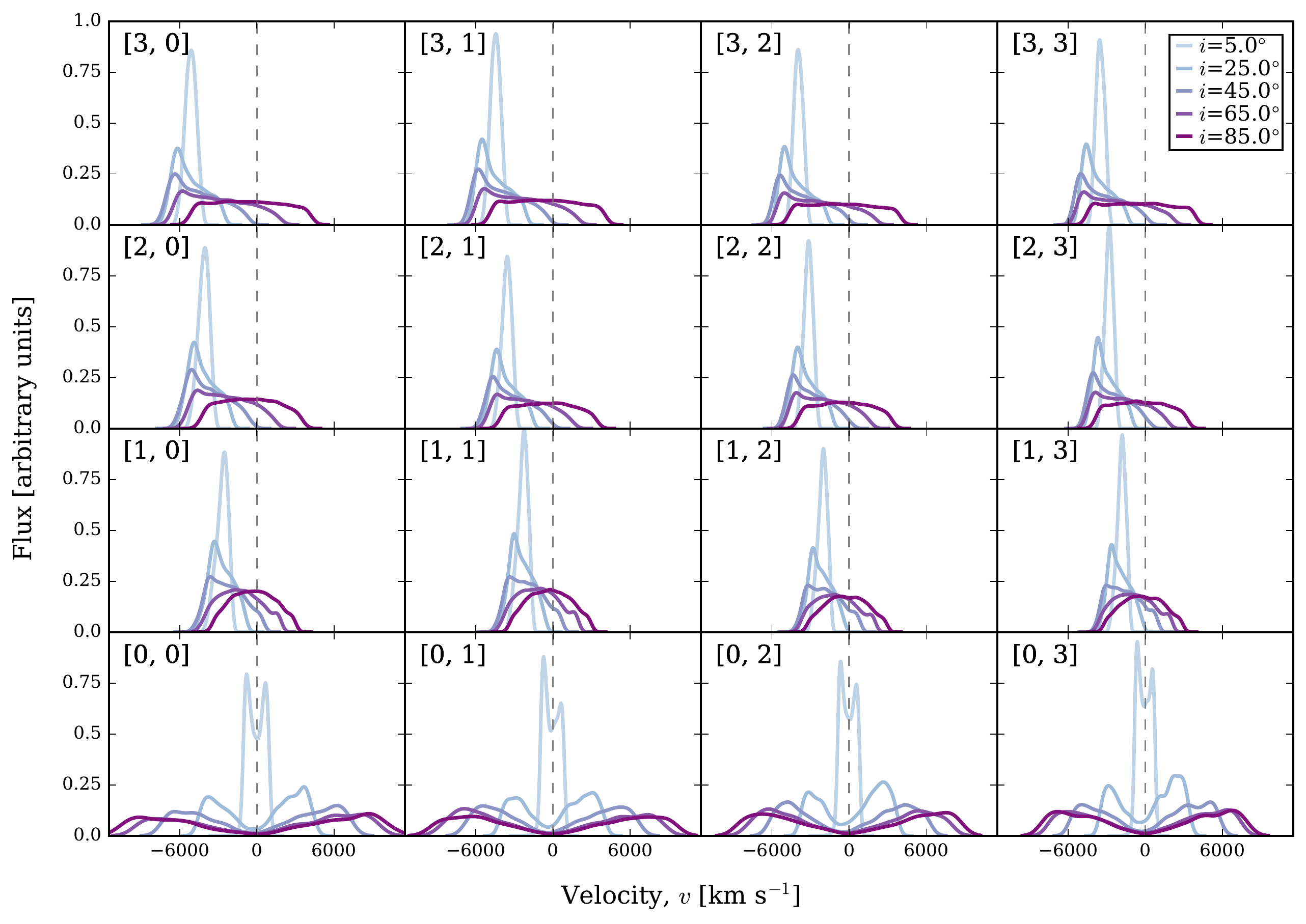}
  \caption{Optically thick intermediate wind with opening angle of $40\degree \text{--} 50\degree$.}
  \label{fig:lpzone_t40-50_xi10}
\end{subfigure}
\caption{Simulated emission line profiles as a function of inclination angle for optically thick wind with $\xi=10^{10}\,$s$^{-1}$. The position of the `wind zone' $[a, b]$ is indicated on the top left of each panel. The dashed line shows the centre of the axis. \label{fig:lpzone_xi10}}
\end{figure*}%
\begin{figure*}[!htbp]
\ContinuedFloat
\centering
\begin{subfigure}[t]{0.83\textwidth}
  \includegraphics[width=\textwidth,keepaspectratio]{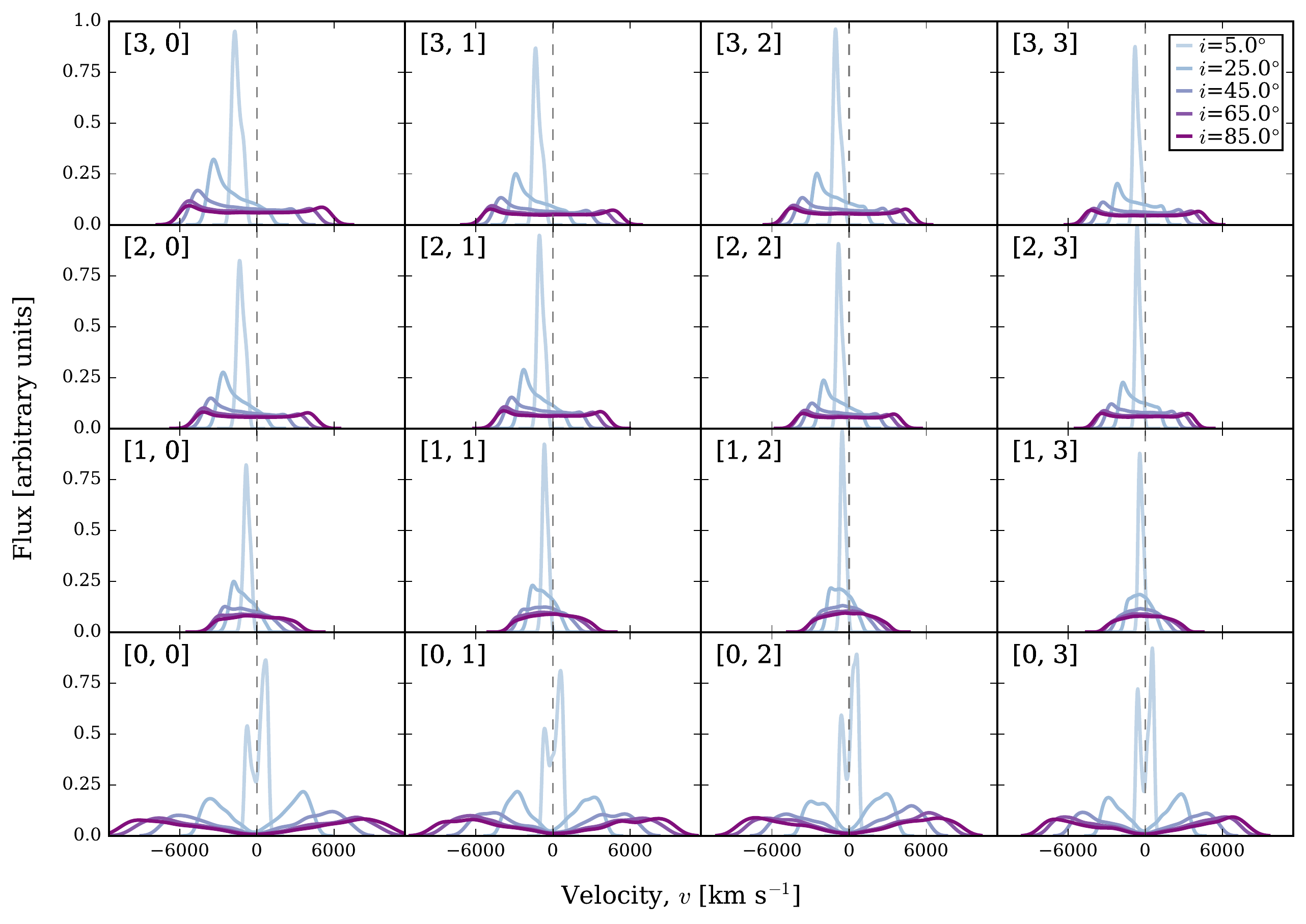}
  \caption{Optically thick equatorial wind with opening angle of $75\degree \text{--} 85\degree$.}
  \label{fig:lpzone_t75-85_xi10}
\end{subfigure}
\caption{--- Continued \label{fig:lpzone_xi10}}
\end{figure*}

\subsection{Time Delay} \label{ssec:rtimedelay}

\begin{figure*}[!htbp]
\centering
\begin{subfigure}[t]{0.33\textwidth}
  \includegraphics[width=\textwidth,keepaspectratio]{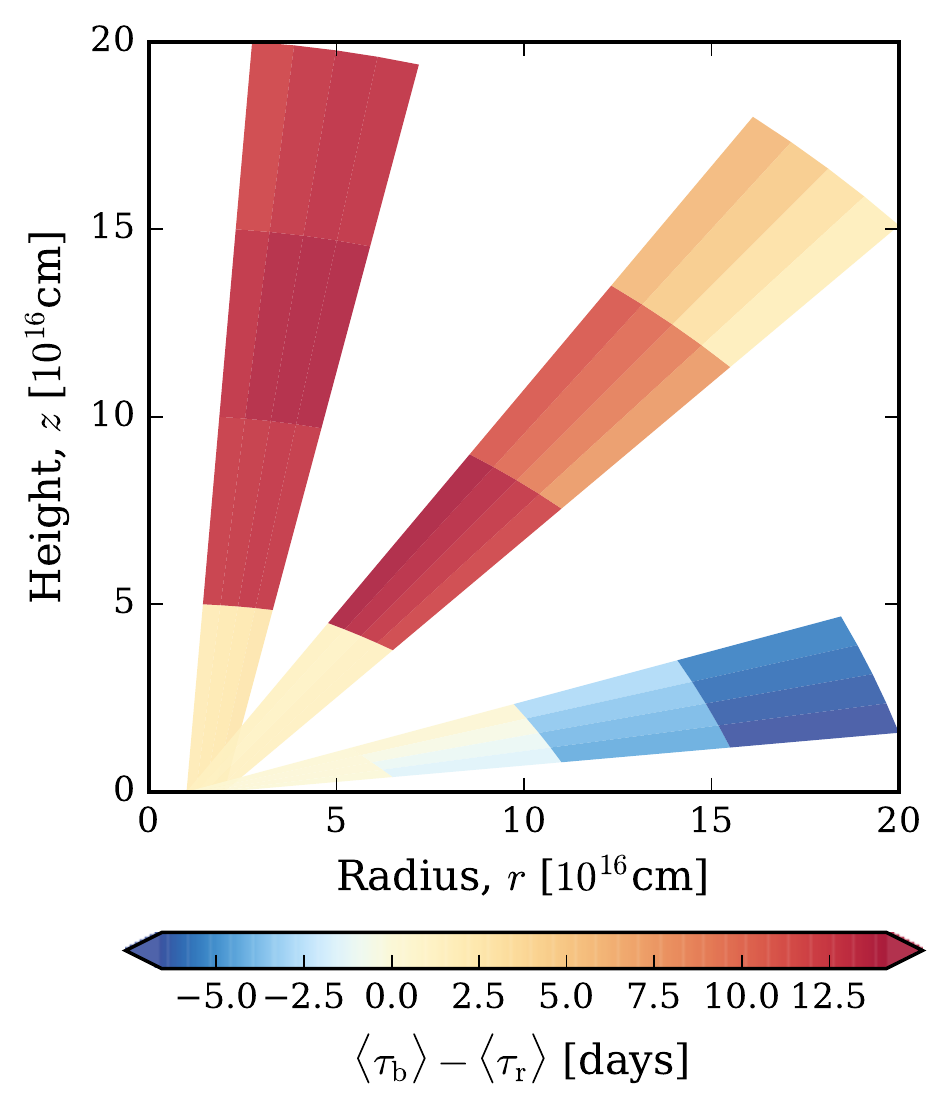}
  \caption{$i=5\degree$}
\end{subfigure}%
\begin{subfigure}[t]{0.33\textwidth}
  \includegraphics[width=\textwidth,keepaspectratio]{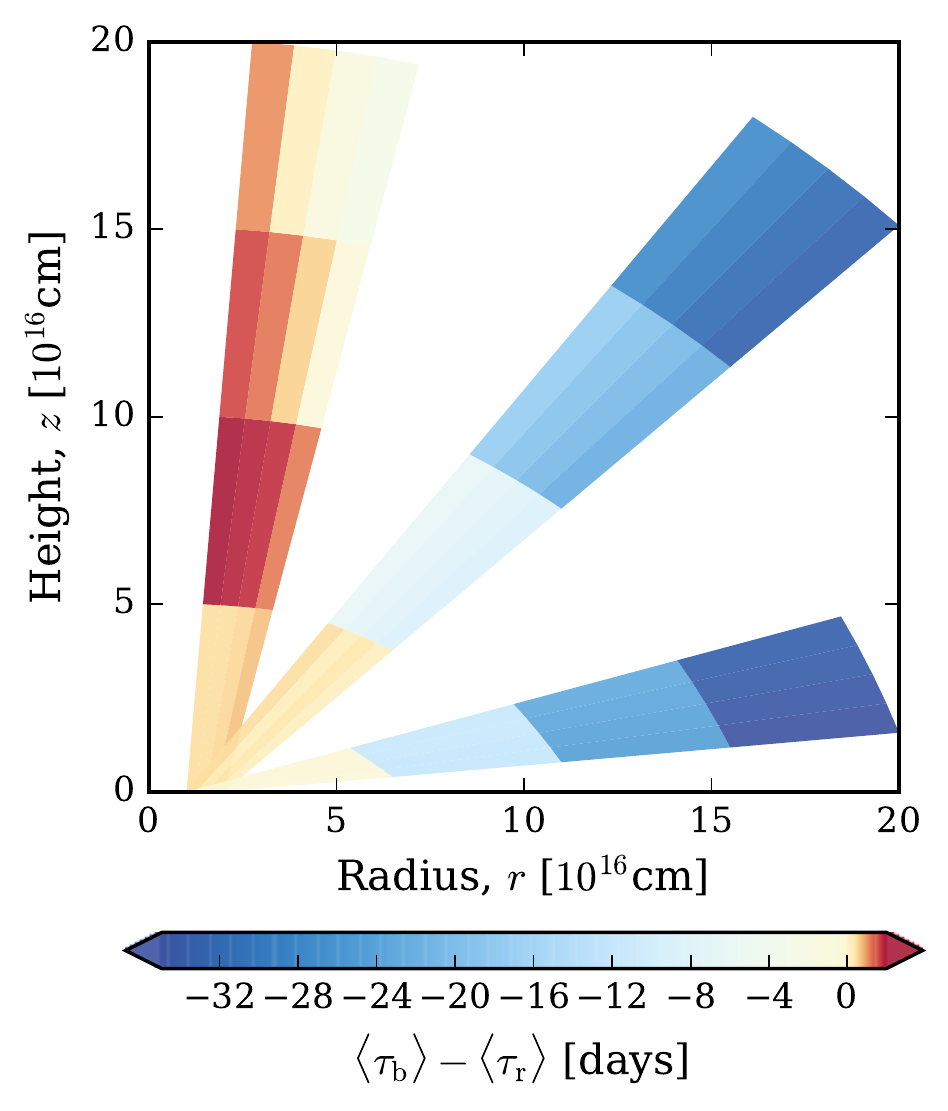}
  \caption{i=$25\degree$}
\end{subfigure}%
\begin{subfigure}[t]{0.33\textwidth}
  \includegraphics[width=\textwidth,keepaspectratio]{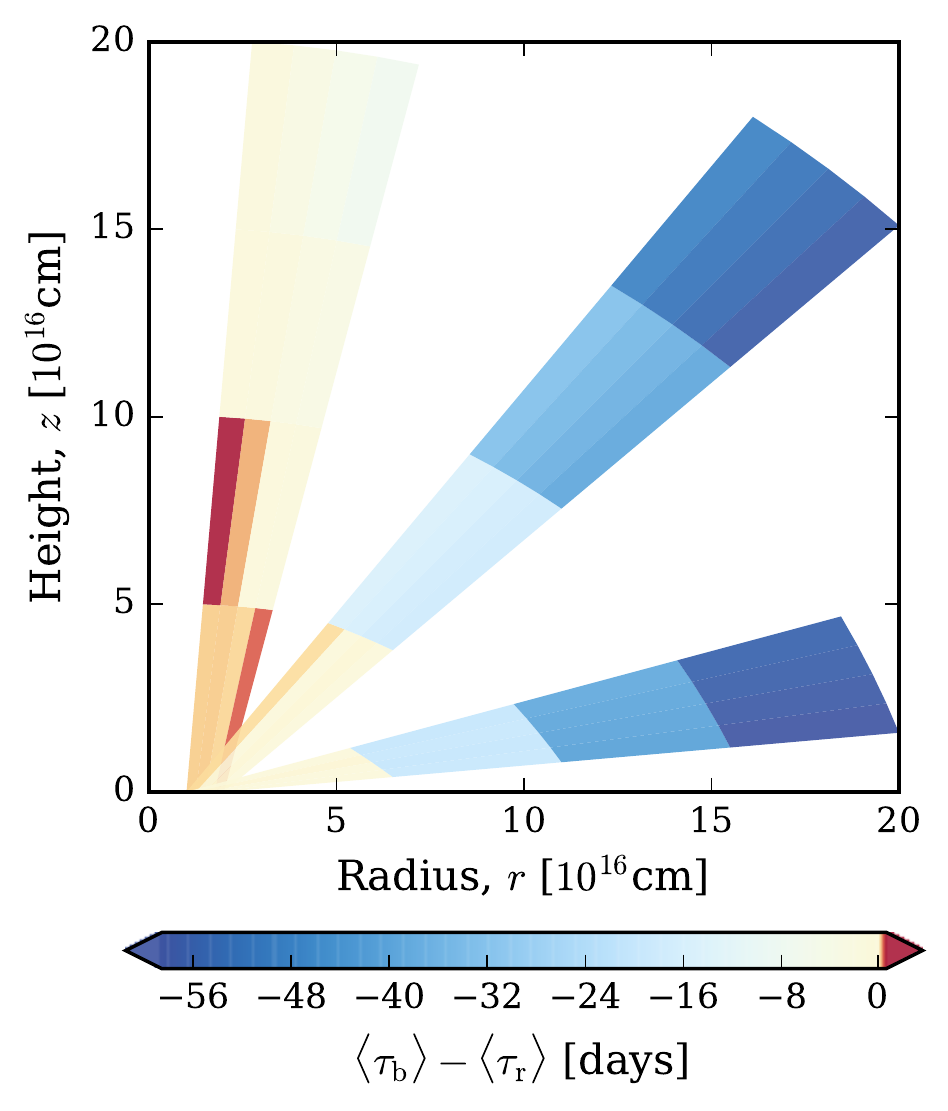}
  \caption{$i=45\degree$}
\end{subfigure}
\\
\begin{subfigure}[t]{0.33\textwidth}
  \includegraphics[width=\textwidth,keepaspectratio]{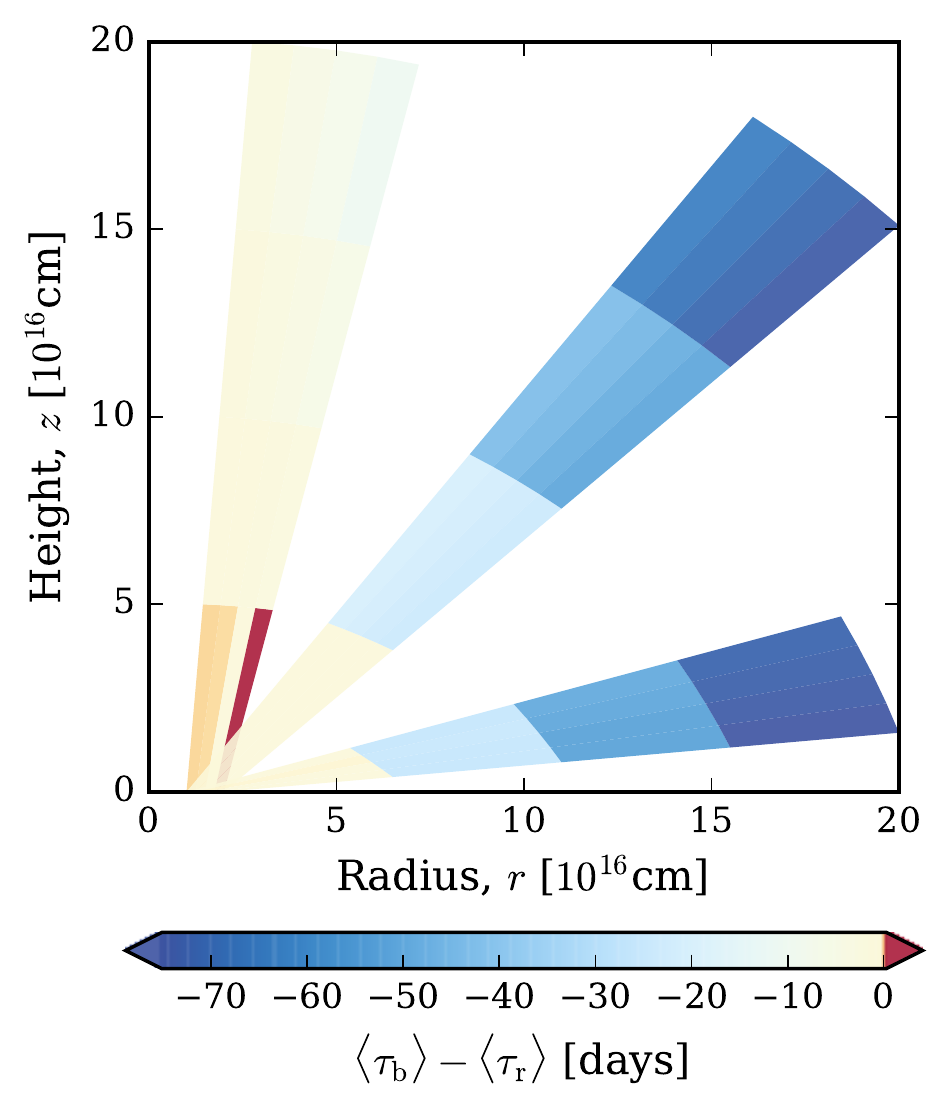}
  \caption{$i=65\degree$}
\end{subfigure}%
\begin{subfigure}[t]{0.33\textwidth}
  \includegraphics[width=\textwidth,keepaspectratio]{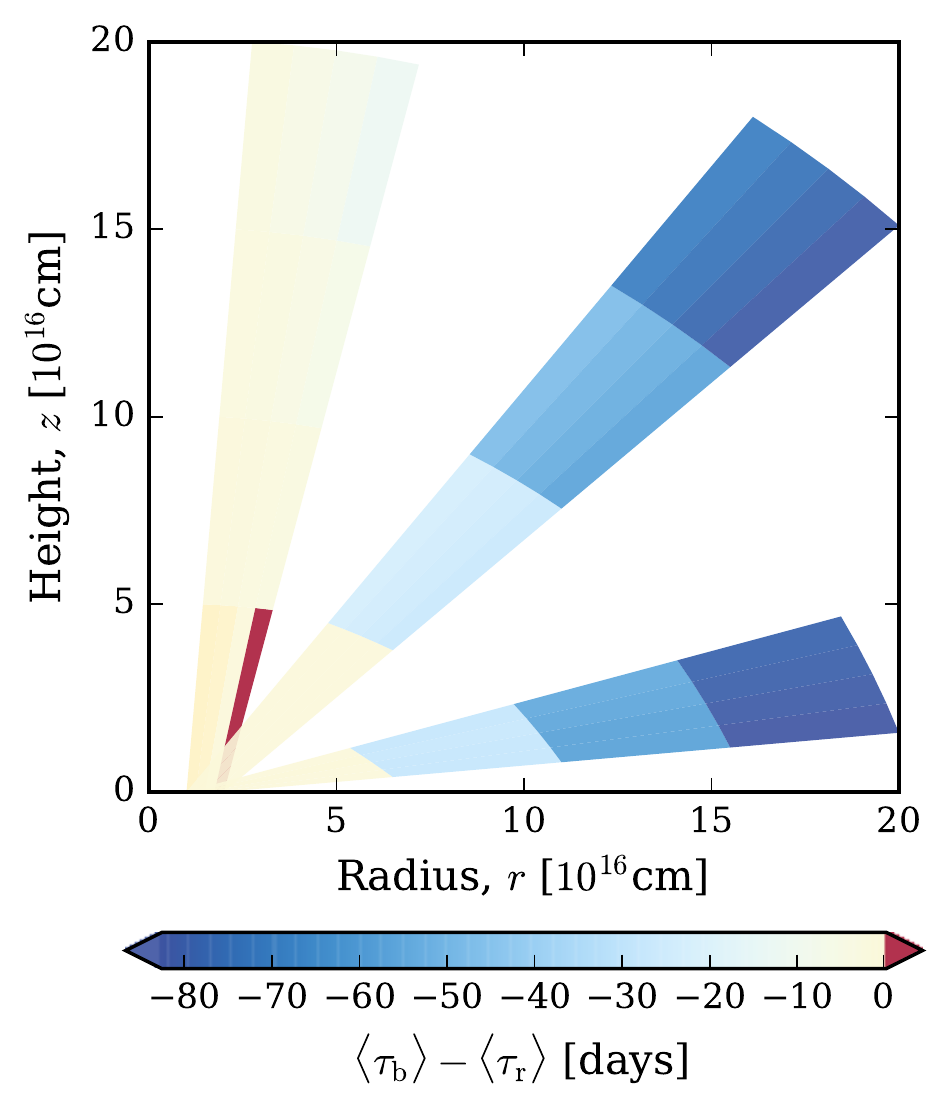}
  \caption{$i=85\degree$}
\end{subfigure}
\caption{Difference between mean time delay of blue and red sides for optically thin wind at various wind opening angles and inclination angles. The narrow winds in each plot: {\em left}: polar, {\em middle}: intermediate, and {\em bottom}: equatorial.\label{fig:diffbrtau_xi0}}
\end{figure*}

\begin{figure*}[!htbp]
\centering
\begin{subfigure}[t]{0.33\textwidth}
  \includegraphics[width=\textwidth,keepaspectratio]{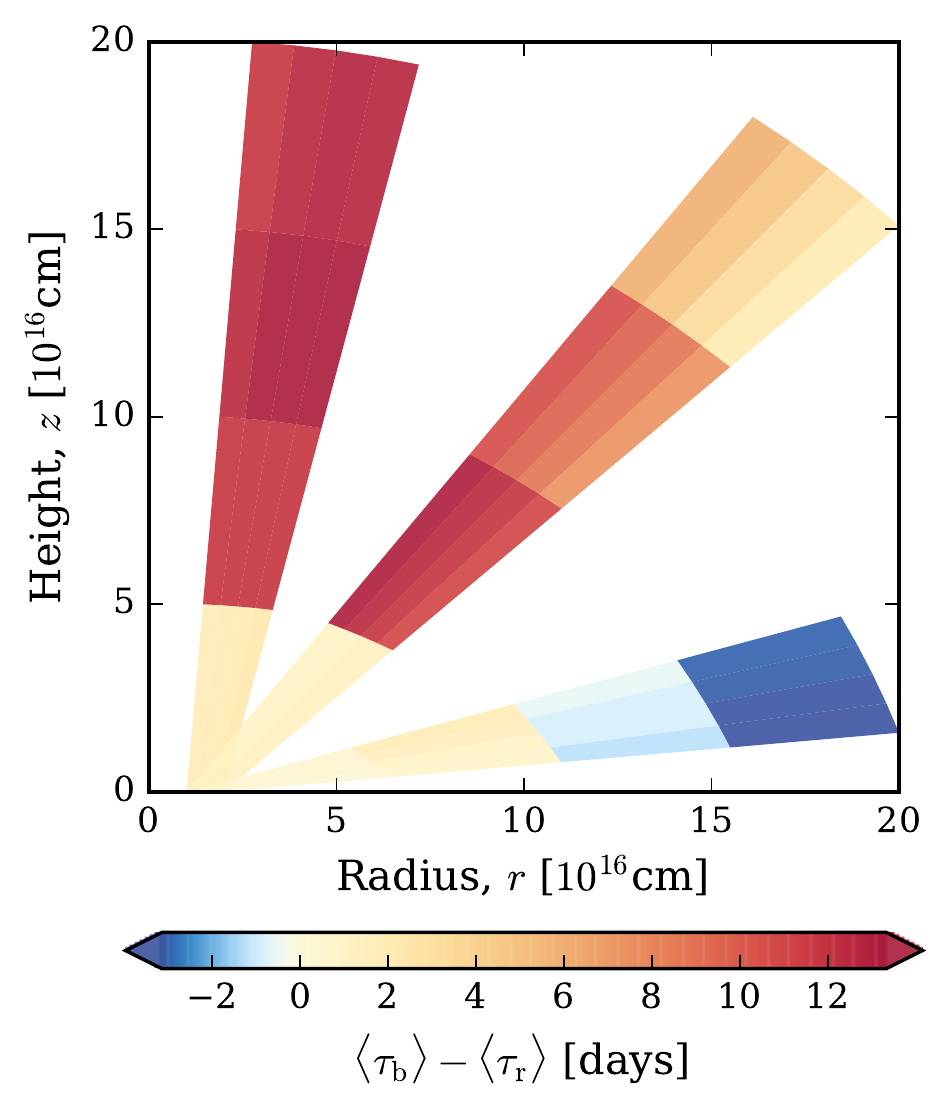}
  \caption{$i=5\degree$}
\end{subfigure}%
\begin{subfigure}[t]{0.33\textwidth}
  \includegraphics[width=\textwidth,keepaspectratio]{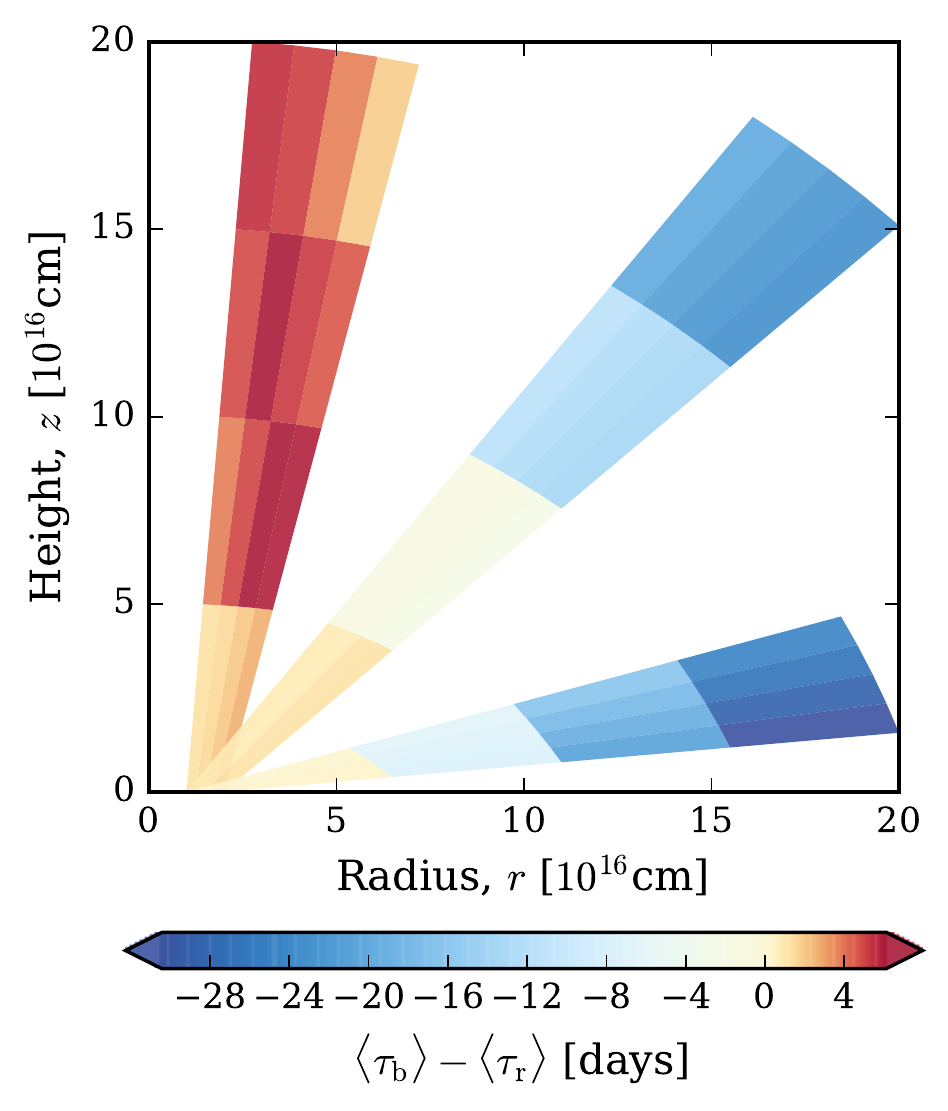}
  \caption{i=$25\degree$}
\end{subfigure}%
\begin{subfigure}[t]{0.33\textwidth}
  \includegraphics[width=\textwidth,keepaspectratio]{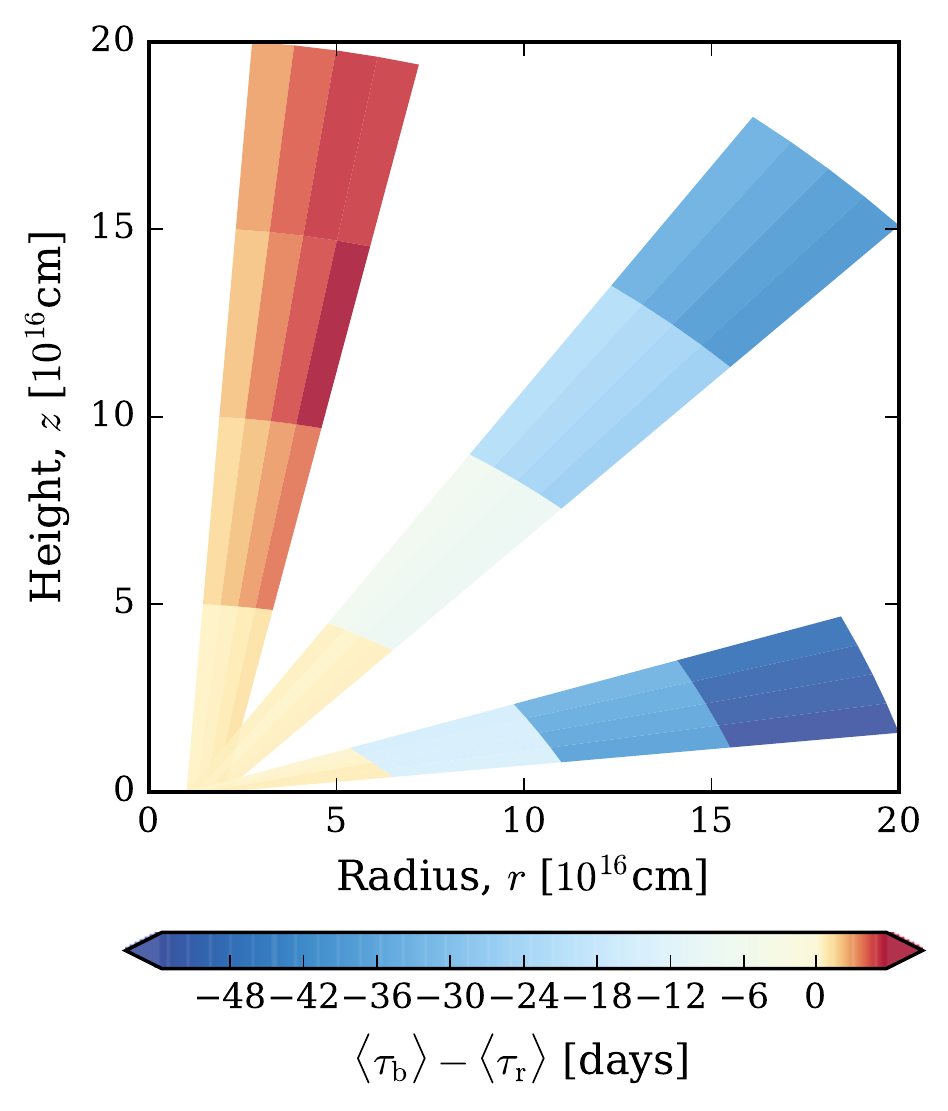}
  \caption{$i=45\degree$}
\end{subfigure}
\\
\begin{subfigure}[t]{0.33\textwidth}
  \includegraphics[width=\textwidth,keepaspectratio]{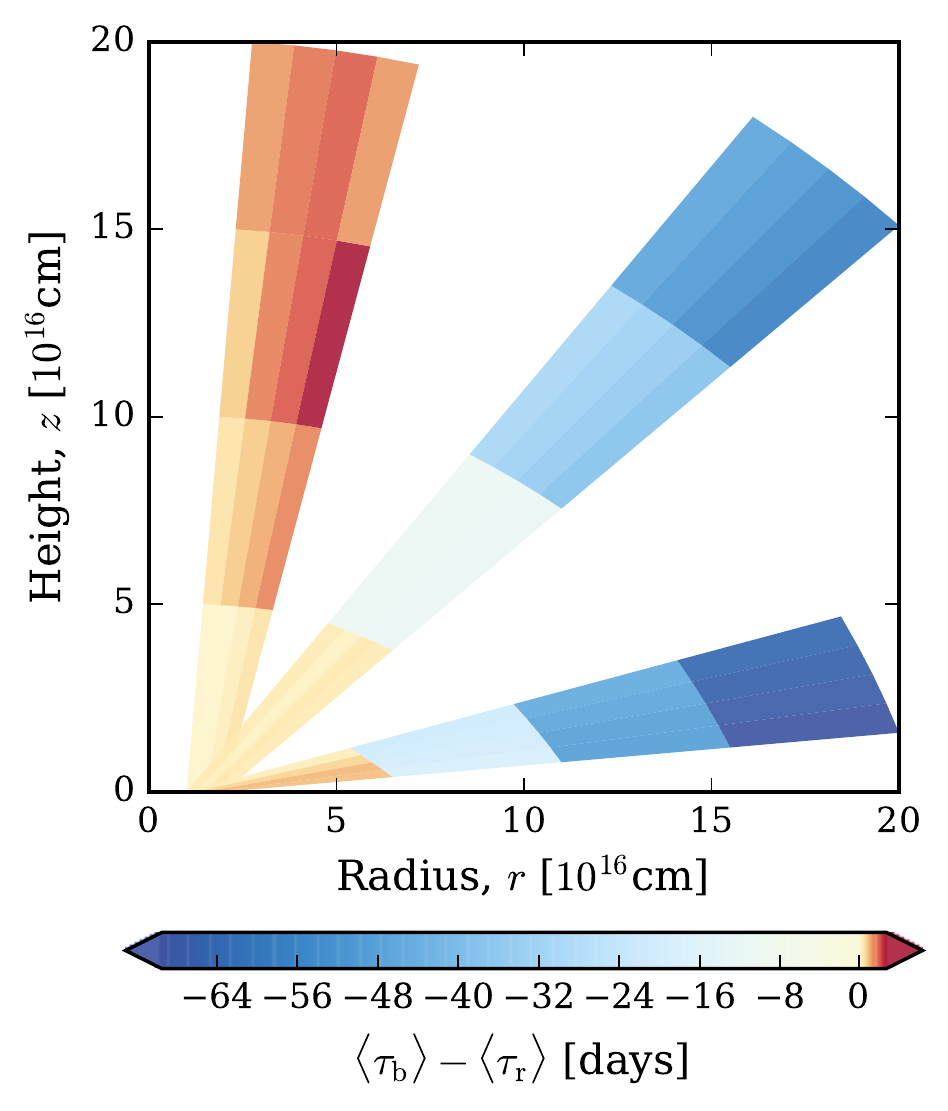}
  \caption{$i=65\degree$}
\end{subfigure}%
\begin{subfigure}[t]{0.33\textwidth}
  \includegraphics[width=\textwidth,keepaspectratio]{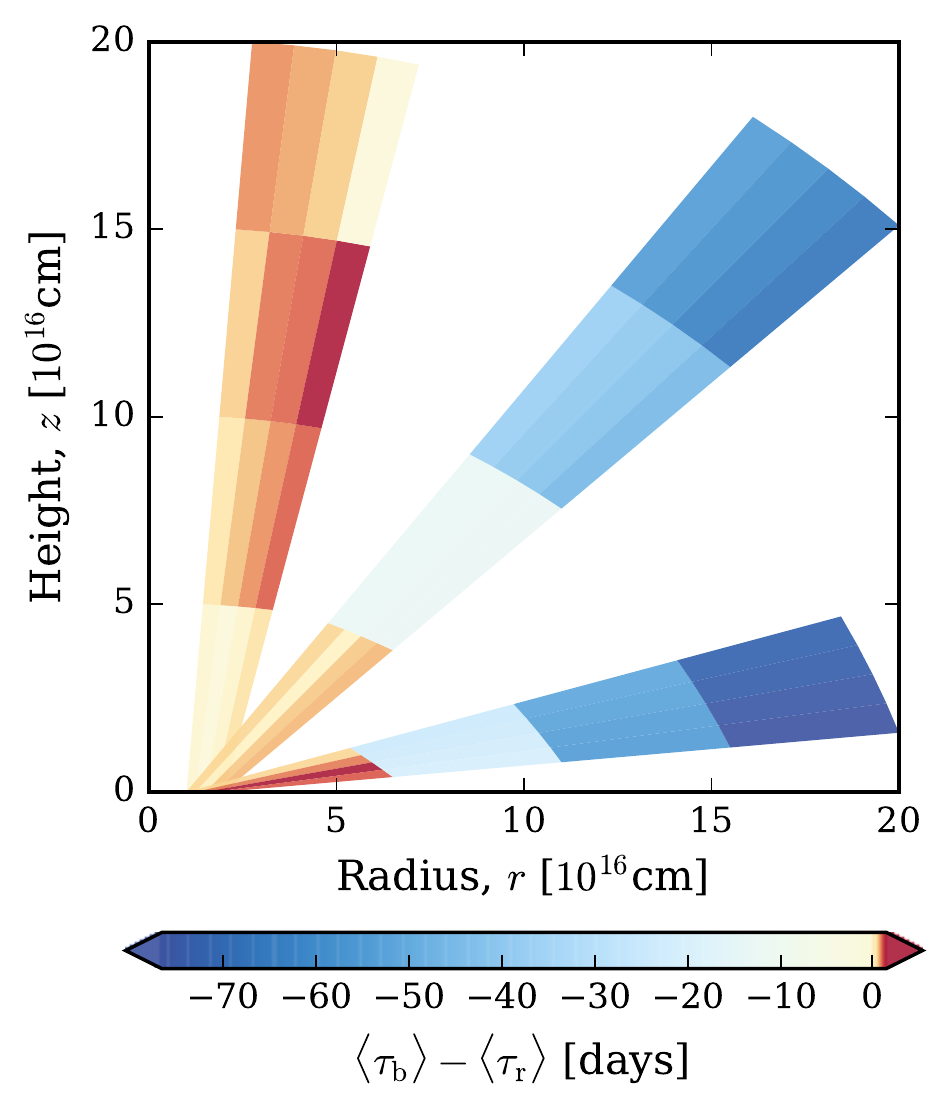}
  \caption{$i=85\degree$}
\end{subfigure}
\caption{Difference between mean time delay of blue and red sides for optically thin wind at various wind opening angles and inclination angles. The narrow winds in each plot: {\em left}: polar, {\em middle}: intermediate, and {\em bottom}: equatorial.\label{fig:diffbrtau_xi10}}
\end{figure*}

Figures~\ref{fig:diffbrtau_xi0} and \ref{fig:diffbrtau_xi10} display the difference between mean time delays of blue and red sides, $\langle\subtext{\tau}{b}\rangle-\langle\subtext{\tau}{r}\rangle$, at varying opening angles of the wind and viewing angles for two conditions of optical depth. The wind is optically thin using $\xi=1\,$s$^{-1}$ in Fig.~\ref{fig:diffbrtau_xi0} and optically thick with $\xi=10^{10}\,$s$^{-1}$ in Fig.~\ref{fig:diffbrtau_xi10}. The mean time delays are colour-coded by their values. A negative difference (blue) implies the blue side response is faster than the red side of the line, and vice versa for positive value (red). Since the time delay is grouped into the blue and red parts from the median line-of-sight velocity of the line profile, a quicker response in the red side does not necessarily correspond to inflowing motion in our model. In the cases where the red side responds quicker than the blue, all or a significant fraction of the particles that make up the red side have a negative line-of-sight velocity (i.e.\ outflowing). Therefore, when the red side responds quicker than the blue side, it simply means that the median time delay to the parts of the wind with larger negative \subtext{v}{los} is longer than the delay to the parts with a more positive \subtext{v}{los}.

Most zones in the polar wind model and a few zones in the intermediate wind model, particularly at small angle of viewing, have positive difference in the mean time delay. This indicates that red side of the line profile responds quicker compared to the blue side. In contrast, the equatorial wind model exhibits shorter lag in the blue side than the red side. The differences are more noticeable at large inclination angles. In all cases, as the inclination angle increases towards edge-on, the difference in mean time delay decreases, i.e., the response in the blue side is becoming faster than the red side.

Changing the optical thickness to a higher value results in smaller time lags of blue side. The mean time delays for the polar wind are generally all positive except for a few zones when the wind is viewed close to edge-on. The intermediate and polar wind show signatures of faster red side response in the zones close to the base in the optically thick wind but not in the optically thin wind.

\section{Discussion} \label{sec:discussion}

Understanding the wind properties that give rise to the diversity of quasar broad line profiles has crucial consequences in inferring the structure of the BLR. When line profiles predicted from given models are compared to the observed emission lines, they can be used to infer the kinematics and dynamics of the line emitting region. In this section, we perform a qualitative analysis on a disk-wind model characterised by a flattened rotating accretion disk and a narrow outflowing helical wind.

\subsection{Wind Opening Angle} \label{ssec:windangle}

Several authors have considered a multitude of combinations of disk and wind BLR components, but in reality, the main distinction between the models is the opening angle of the wind. Our simulation with the intermediate wind opening angle is similar to that proposed by \citet{Elvis:2000,Elvis:2004}, but without the vertical outflow, which is imposed to justify narrow absorption lines. Our equatorial wind model also resembles \citet{Murray+:1995} model, though their model has a wider opening angle with the wind closer to the base of the disk.

Based on Figs.~\ref{fig:lpzone_xi0} and \ref{fig:lpzone_xi10}, an equatorial wind model, like that of \citet{Murray+:1995}, produces more symmetric and less blueshifted emission line profiles. As the wind opening angle is shifted towards polar, the profiles tend to become more asymmetric with higher blueshifts. Since the wind is outflowing, this indicates that the most of the wind is travelling towards the observer and smaller fraction is projected in the opposite direction. The increase in blueshift implies that the majority of the wind velocity is aligned with the line-of-sight of the observer.

\subsection{Inclination Angle} \label{ssec:iangle}

The degree of line asymmetry tends to increase with decreasing inclination. \citet{Chajet+Hall:2013,Chajet+Hall:2017} attained similar findings using a model combining the improvised accretion disk-wind model of \citet{Murray+Chiang:1997} and the magnetohydrodynamic model of \citet{Emmering+:1992}. We find that the line profiles are symmetric when viewed near edge-on and asymmetric near face-on. This trend is a simple consequence of the outflowing wind. For low inclination objects, the poloidal or outflowing velocity dominates the observed line-of-sight velocity. Meanwhile, in high-inclination-angle objects, the line-of-sight velocity is dominated by the rotational velocity component, which has roughly equal approaching and receding velocity components.

The line profiles also tend to be more blueshifted for smaller inclination angle than those seen at high inclination. Following a similar explanation, the direction of the outflowing wind determines the strength of the blueshift. Viewing the object face-on, the projected wind velocity is towards the line-of-sight of the observer. Assuming that the far side of the emission is obscured by the optically thick accretion disk, this yields an overall bluewards shift of the line. In the optically thick wind, the emission line gradually becomes more blueshifted as the line-of-sight approaches and intersects the opening wind region within \subtext{\theta}{min} and \subtext{\theta}{max} (Fig.~\ref{fig:lpzone_xi10}). When the inclination angle passes the emitting region and moves closer towards edge-on, the lines become less blueshifted since the line-of-sight is now travelling away from the wind. In addition, the Keplerian rotation component significantly contributes to the projected velocity and hence, the increase in line width. These findings are consistent with the results from \citet{Chajet+Hall:2013,Chajet+Hall:2017} line profile modelling.

\subsection{Wind Zone Position} \label{ssec:windzonepos}

Depending on the spatial location of the line emission region in the wind, there will be a variation in the physical properties of the emission lines. The width of the line profile is broader and approximately symmetric near the base of the streamline. This is expected since the rotational velocity or Keplerian motion is dominant close to the surface of the accretion disk. As the wind travels outwards in poloidal distance, the rotational velocity starts to decrease, while the poloidal component of the velocity gradually increases. This contributes to the  narrowing of the line profile for regions further out until at a certain point where the poloidal velocity is significant, and consequently broadens the line profile. This transition is particularly strong in more equatorial winds as demonstrated in Figs.~\ref{fig:lpzone_t40-50_xi0}\text{--}\ref{fig:lpzone_t75-85_xi0} and Figs~\ref{fig:lpzone_t40-50_xi10}\text{--}\ref{fig:lpzone_t75-85_xi10}.

Earlier studies have shown that the shape of the emission line profile depends on the ionisation level of the line. HILs, like \ion{C}{iv}, are often broader in comparison to LILs, like \ion{Mg}{ii} and H$\beta$ \citep[e.g.,][]{Osterbrock+Shuder:1982,Mathews+Wampler:1985}. The distinct line shapes indicate different physical conditions in the line emitting region and thus, it is possible to infer the relative spatial location of the line in the outflowing wind. It is expected that the LILs will lie close to the base of the wind further from the black hole, in a region of higher density \citep{Ruff+:2012}, while the HILs will lie closer to the ionising source, and higher in the wind, reflecting both the rotational and the poloidal wind components. This explains the reliability of the LILs in black hole estimation \citep[e.g.,][]{McLure+Jarvis:2002,Shen+:2008,Rafiee+Hall:2011,Mejia-Restrepo+:2016} and the observed velocity offset of the HILs with the systemic velocity of the system compared to that of the LILs \citep{Hewett+Wild:2010}.

The relative positions of the HILs and LILs agrees with the blueshifting trend in our model. The blueward shift in the line is higher for `wind zones' close to the ionising source relative to zones further away but near the base of the wind. This is consistent with the observed blueshift of HILs with respect to LILs \citep[e.g.,][]{Gaskell:1982}. \ion{C}{iv} HILs that display large blueshifts of $>2000\,\kms$ tend to be dominated by non-virial motion \citep{Coatman+:2016}. In our model, the blueshifting is most prominent for polar wind opening angle due to the reason mentioned in previous section. The emission lines in the equatorial wind only show slight blueshifts with larger poloidal distance.

\subsection{Optical Depth} \label{ssec:opticaldepth}

In the optically thin wind situation, most of the simulated lines presented are double-peaked, which are rarely observed \citep{Eracleous+Halpern:1994,Eracleous+Halpern:2003,Strateva+:2003}. The profiles are generally single-peaked for viewing angle close to face-on. However, `wind zones' that are close to the accretion disk and for polar outflowing wind opening angle, always show double-peaked profiles.

As suggested by \citet{Chiang+Murray:1996,Murray+Chiang:1997}, one of the determining elements for a line profile to be single-peaked is a larger velocity gradient in the poloidal compared to the rotational, $|(\diff v_{l}/\diff r)/(\diff v_{\phi}/\diff r)| \gtrsim 1$ since the light is more readily transmitted in a radial direction. In our model, this ratio is always less than 1 in all `wind zones' for a polar outflowing wind. Consequently, the profiles remain double-peaked regardless of the optical thickness of the wind. In the intermediate and equatorial wind models, the rotational shear at regions close to the accretion disk is also higher than the radial shear. Hence the double-peaked line profiles. However, the radial shear increases with poloidal distance and the photons are more likely to escape radially along the line-of-sight, which results in the observed single-peaked lines.

\subsection{Time Delay} \label{ssec:dtimedelay}

The response of the BEL flux to changes in the ionising continuum flux reflects the geometric configuration of the BLR. The side closest to the observer is seen to vary with changes in the continuum flux earlier \citep{Gaskell:2009}. Generally, a shorter lag in the red side of the line is associated with inflow motion, while a shorter blue side response is related to outflow motion. Many objects monitored in velocity-resolved RM are detected to have the red side of the line profile leading the blue side and hence, disfavouring outflow models and supporting inflow cases \citep[e.g.,][]{Gaskell:1988,Koratkar+Gaskell:1989,Crenshaw+Blackwell:1990,Korista+:1995,Ulrich+Horne:1996}. However, using the spherical disk-wind model of \citet{Murray+:1995}, \citet{Chiang+Murray:1996} demonstrated that it is possible to attain earlier response in the red side of the line by taking into account radiative transfer effects due to the radial and rotational components of the velocity.

Our kinematical narrow wind model is also able to recreate the shorter time delay in the red or blue side of the line for both optically thin and thick winds. The time lag in the red side is quicker than the blue side for polar and intermediate wind opening angles, especially when the viewing angle is close to face-on. This is in accordance with Type~1 objects seen at low inclination angle. A strong indication of outflowing winds with shorter lag in the blue side is exhibited in equatorial winds at high viewing angle.

\subsection{Parameter Sensitivities} \label{ssec:paramsensitivity}

The results presented assume the model parameters described in Section~\ref{ssec:kinematics}. However, despite the exact values chosen for this analysis, the underlying trends in line widths and blueshifts at varying viewing angles and outflowing wind should remain true over a large range of parameters. In order to test this, we inspect how changing the values of some parameters affects the shape of the line profile.

We found that changing the source function power law exponent, $\beta$, in the radiative transfer equation only slightly changes the line profile width and asymmetry. At regions far from the surface of the disk, the increase in $\beta$ yields narrower emission lines in the equatorial wind model. This trend is in agreement with the \citet{Murray+Chiang:1997} studies. For intermediate wind angles, the lines are narrower with less flux on the blue wing and slightly more on the red wing. However, in all wind models, zones near the base display broader lines with higher $\beta$. The rest of the zones for a polar wind also show line profiles broader on the red side and slightly narrower on the blue side. This trend is likely because zones have smaller poloidal velocity shear than rotational shear with $|(\diff v_{l}/\diff r)/(\diff v_{\phi}/\diff r)| < 1$.

The widths of the line profiles were found to be broader with increasing black hole mass. They scale roughly $\propto \subtext{M}{BH}^{1/2}$ as expected from virial motion, assuming other parameters are kept constant. The rate of wind acceleration can be regulated by the quantity $\alpha$. For a large $\alpha$ value, the poloidal velocity near the base of the accretion disk starts slow initially but gradually increases at around $R_{v}$. On the other hand, a smaller $\alpha$ implies a faster initial poloidal velocity.

The density of the wind is dependent on the parameters $\lambda$ and \subtext{\dot{M}}{wind}. A negative mass-loss rate exponential leads to a decreasing local mass-loss rate as the radius increases. By lowering the mass-loss rate of the wind, the density decreases and causes a change in ionisation states. In all scenarios, the qualitative trends in the line profile shapes are fairly similar.

\subsection{Caveats} \label{ssec:caveats}

Several caveats are noteworthy in our modelling. The values of $\xi$ in the optical depth function were selected to represent an optically thin or thick wind. Although these conditions are shown to suppress the double peaks to a single peak, a more sensible value obtained through photoionisation simulations should further clarify the effects of optical depth on the emission lines. There is also a possibility for radiative transfer in multiple scattering surfaces \citep{Rybicki+Hummer:1978}.

Though our modelling assumes that the line profiles are generated locally within each zone, in a realistic context, variations in density can induce significant fluctuations in the ionising state. The density ranges at different opening angles are $\sim 10^{-17} \text{--} 10^{-13}\,$g\,cm$^{-3}$ (Hydrogen number density, $\subtext{n}{H} \sim 6 \times 10^{6 \text{--}10}\,$cm$^{-3}$) for polar wind, $\sim 10^{-18} \text{--} 10^{-13}\,$g\,cm$^{-3}$ ($\subtext{n}{H} \sim 6 \times 10^{5 \text{--} 10}\,$cm$^{-3}$) for intermediate wind, and $\sim 10^{-18} \text{--} 10^{-12}\,$g\,cm$^{-3}$ ($\subtext{n}{H} \sim 6 \times 10^{5 \text{--} 11}\,$cm$^{-3}$) for equatorial wind. Huge variations in densities are seen for $[0, b]$ `wind zones' that are close to the base of the accretion disk.  The difference can be up to three orders of magnitude $\sim 10^{-16} \text{--} 10^{-13}\,$g\,cm$^{-3}$ for polar winds and up to four orders of magnitude for intermediate and equatorial winds with ranges of $\sim 10^{-17} \text{--} 10^{-13}\,$g\,cm$^{-3}$ and $\sim 10^{-16} \text{--} 10^{-12}\,$g\,cm$^{-3}$ respectively. The densities vary approximately an order of magnitude for the rest of the `wind zones'. Although most emission lines are effective at emitting over a large range of densities \citep{Korista+:1997}, ionisation changes over these density ranges could dramatically affect the emergent line profiles. As the wind travels further in poloidal distance, the changes in density are lesser. The model can be enhanced by taking into account the complexity of the BLR structure, such as incorporating photoionisation and line driving mechanisms.

\subsection{Comparison with Other Studies} \label{ssec:comparison}

Different models of quasar wind geometries have been tested against the observations by a number of authors. \citet[][and references therein]{Marin+:2015} considered different models to explain the observed polarisation dichotomy between Type~1 and 2 AGN. This dichotomy shows Type~1 quasars typically have polarisation parallel to the system axis, while for Type~2, the polarisation is perpendicular to the axis. They found that a two-phased outflowing wind with a bending angle of $45\degree$ is able to explain this dichotomy. However, their predicted polarisations depend quite strongly on the clumpiness of the wind and the line-of-sight. Similarly \citet{Young+:2007} modelled the structures observed in polarized light of the BAL quasar PG1700+518, and found the general geometric structure of both the BLR wind and the electron scattering wind responsible for the polarisation consistent with an outflowing wind, with a poloidal launching angle.

Other authors are challenging the idea that quasars follow a simple unification model, depending only on black hole mass and accretion rate \citep[for example,][and references therein]{Matthews+:2017,DiPompeo+:2017}. In \citet{DiPompeo+:2017}, evidence is presented for evolutionary differences affecting the amount of obscuration in quasars and \citet{Matthews+:2017} concluded that BAL quasars are viewed from similar angles to non-BAL quasars, or geometric unification cannot explain the fraction of BALs in quasar samples.

Detailed hydrodynamic simulations of a line-driven wind have illustrated the possibility of over-ionising the wind, leading to a reduction in efficiency of the line-driving. This illustrates the complexity of the BLR region demanded by observations, which require shielding of the outflowing winds \citep{Proga+:2000,Higginbottom+:2014}. These different models and observations will be considered in detail in a subsequent paper, \citet{Yong+:2017}.

\section{Summary} \label{sec:summary}

We have explored the properties of BELs, specifically the widths and velocity offsets, using a dynamical disk-wind model with radiative transfer in the Sobolev limit. The effect of orientation for narrow angle of outflowing wind is analysed. We have considered several factors that contribute to the different BEL features, which include inclination angle, angle of outflow, and position of the emission region in the wind.

When viewed face-on, the emission line profile is asymmetric and narrow. The profile is blueshifted since the wind is approaching the observer. The blueshift increases as the inclination angle moves toward the opening angle of the wind and decreases as the viewing angle recedes from the emitting wind. The profile is symmetric and broader as the viewing angle approaches edge-on. At wind regions close to the accretion disk surface, the emission line is symmetric with broader width than a line emitted further out. The emission line profile has a larger blueshift with increasing vertical distance of the wind from the central ionising source, particularly for polar wind opening angle. The blueshifting effect decreases as the angle of outflowing wind tends towards equatorial.

By taking into account the correction for optical depth, single-peaked emission lines are formed for the intermediate and equatorial wind. Due to the relatively small poloidal velocity gradient compared to the rotational shear found in the polar wind, the lines are double-peaked even after applying the optical depth correction.

We have also demonstrated that an ouflowing wind model is capable of generating a shorter lag in the red or blue sides of the line. This suggests that outflowing winds are not ruled out when the red side of the emission line responds on a quicker time frame than the blue side, as has previously been implied.

\begin{acknowledgements}
We thank the anonymous referee for valuable suggestions on the manuscript. NFB thanks the STFC for support under Ernest Rutherford Grant ST/M003914/1.
\end{acknowledgements}


\end{document}